\documentclass[aps,preprint,nofootinbib,groupedaddress]{revtex4}
\usepackage[dvips]{graphicx}
\usepackage{amsmath,amssymb,graphicx,subfigure}
\newcommand{\fm}{{\textrm fm}}
\newcommand{\GeV}{\textrm{GeV}}
\newcommand{\MeV}{\textrm{MeV}}
\newcommand{\be}{\begin{eqnarray}}
\newcommand{\ee}{\end{eqnarray}}

\renewcommand{\sc}{\slashchar}
\newcommand{\ds}[1]{
#1{\hskip-2.0mm}/
}
\def\slashchar#1{\setbox0=\hbox{$#1$}           
   \dimen0=\wd0                                 
   \setbox1=\hbox{/} \dimen1=\wd1               
   \ifdim\dimen0>\dimen1                        
      \rlap{\hbox to \dimen0{\hfil/\hfil}}      
      #1                                        
   \else                                        
      \rlap{\hbox to \dimen1{\hfil$#1$\hfil}}   
      /                                         
   \fi}                                         %

\begin{document}

\title{Exploring the Chiral Regime of QCD in the Interacting Instanton Liquid Model}
\author{M.~Cristoforetti, P. Faccioli and M.C. Traini}
\affiliation{Dipartimento di Fisica and I.N.F.N., Universit\`a degli Studi di Trento, Via Sommarive 15, Povo (Trento) 38050 Italy.}
\affiliation{European Centre for Theoretical Studies in Nuclear Physics and Related Areas, Strada delle Tabarelle 286, Villazzano (Trento), I-38050 Italy.}
\author{J.W.~Negele}
\affiliation{ Center for Theoretical Physics
Massachusetts Institute of Technology, NE25-4079, 
77 Massachusetts Ave, Cambridge, MA 02139-4307, USA.}


\begin{abstract}

The Interacting Instanton Liquid Model is used to explore the role of instanton induced dynamics in hadron structure. To support the validity of this model in the chiral regime, the quark mass dependencies of several properties are shown to agree with chiral perturbation theory, including the density of eigenmodes of the Dirac operator and the masses of the pion and nucleon. A quark mass $m^*$ = 80 MeV emerging naturally from the model is shown to specify the mass scale above which the fermion determinant is suppressed, the zero modes become subdominant, and the density of quasi-zero modes become independent of the quark mass.
\end{abstract}  

\maketitle

\section{Introduction}

The goal of this work is to gain insight into the mechanism by which the structure of hadrons arises from QCD. Although lattice field theory provides a powerful tool for solving nonperturbative QCD and is now beginning to successfully  calculate experimental observables  with dynamical quarks in the chiral regime, the physical mechanism by which hadron structure arises and the dominant degrees of freedom are not directly evident.  Hence, in this work, we explore the complementary insight that can be obtained from a model that focuses on what we believe to be the dominant degrees of freedom relevant to hadron structure.  In nonperturbative QCD, the way in which gluons interact with quarks depends dramatically on the quark mass.  In the limit of large mass, heavy quarks move adiabatically in a flux tube potential, whereas in the limit of light mass, a variety of evidence suggests that the instanton-induced 't Hooft  quark interaction plays an important role and provides the mechanism for spontaneous chiral symmety  breaking.  To understand the structure of hadrons containing light up and down quarks, we therefore seek to explore the role of  instantons and their associated zero modes in hadron structure, and do so in the context of the interacting instanton liquid model (IILM).  

In the IILM, the QCD path integral over all gluons is replaced by an effective theory in which instantons are the effective degrees of freedom, and the gauge fields of the theory are those generated by integrating over the positions, color orientations, and sizes of instantons.  In the context of this model, we would like to understand the mass range in which instanton mediated chiral dynamics is manifested and the extent to which it is described by chiral perturbation theory.
In particular, we would like to know the mass scale or scales at which the continuum fermion determinant is suppressed, at which zero modes become subdominant, and at which the density of quasi zero modes becomes independent of quark mass.

In the present paper, we set-up the formalism to use the IILM~\cite{IILM} to address these questions. 
The instanton picture (for recent reviews see \cite{rev,diakonov1}) was originally introduced as a model for the QCD vacuum based on semi-classical arguments~\cite{shuryak82}. It was shown that instantons lead to spontaneous chiral symmetry breaking by introducing strong non-pertrubative correlations between fermionic zero-modes localized around the instanton positions. Specific features of the instanton picture have been observed in a number of lattice studies \cite{Negele:1998ev,latticeILM1,latticeILM2,latticeILM3,latticeILM4} and there is evidence that chiral symmetry breaking is correlated with smooth lumps of topological charge, whose profile is consistent with  that of singular-gauge instantons~\cite{latticeILMlumps1,latticeILMlumps2}. In addition, instanton-induced correlations in hadrons have been studied in a number of phenomenological model calculations, where it was shown that the  Instanton Liquid Model, (ILM), provides a  good description of the mass and the electro-weak structure of pions, nucleons and hyperons~\cite{instcorr,RILMm,RILMb,emff1,emff2,emff3,pion1,delta12,diquark,diakonov2}. 

Before trusting the IILM to provide useful insight into the role of instantons and their associated zero modes, it is important to verify several essential properties.   One key issue is to verify the chiral behavior of the spectrum of the Dirac operator.
 By analyzing the dependence on the quark mass of the density of eigenvalues of the Dirac operator, $\rho(\lambda)$. we show that, in the chiral limit, the IILM results are consistent with the well-known chiral perturbation theory ($\chi$pt)  result~\cite{smilga}: 
\be
\lim_{\lambda\to 0}\lim_{m_q\to 0} \rho(\lambda)= \textrm{Const} + O(\lambda^2)\qquad (N_f=2).
\ee
In addition, we check that at small but-finite quark masses, the IILM generates mass corrections to the Dirac spectrum that are consistent with those predicted by $\chi$pt. To do so, we show that from  the structure of the IILM Dirac spectrum computed at different quark masses, one can predict the power-law infrared divergence in the quark mass of the scalar three-point correlator: 
\be
K^{abc} &\equiv& \int \textrm{d}^4x\textrm{d}^4y\textrm{d}^4z\langle 0|S^a(x)S^b(y)S^c(z)|0\rangle\propto \frac{1}{m_q},\\
S^a(x) &=&\bar{q}(z)\tau^a q(x).
\ee
This result agrees with the prediction of $\chi$pt~\cite{smilga}:
\be
K^{abc} = \frac{B^3(N_f^2-12)}{64\pi^2m_{\pi}^2 N_f}d^{abc} + O(p^4)\propto \frac{1}{m_q} .
\ee

A second important test is verifying  that the IILM effective theory can be used to calculate hadron masses and that these masses have the dependence on the quark mass expected from $\chi$pt. Since correlation functions in the nucleon and pion channels receive contributions from a single instanton  whereas in the $\rho$ and $\Delta$ channels  the leading contribution occurs for two instantons, the nucleon and pion masses depend on the instanton density to first order while the $\rho$ and $\Delta$ are only affected in second order.  Hence, we focus here on the nucleon and pion masses.  Since we have replaced the functional integral over all gluon fields by a sum over instanton fields, a key task is to verify that the effective theory is an adequate approximation to the full functional by showing that the correlation function decays at large Euclidean time like a pure exponential, as it must in a field theory with a transfer matrix, and that spontaneous chiral symmetry breaking occurs as it would in the full field theory. In the case of the pion, we demonstrate very clear asymptotic decay as a single exponential and use the slope to obtain an accurate measurement of the pion mass at each quark mass. These data not only have a quark mass dependence consistent with the form expected from chiral perturbation theory, but we show that they yield low energy constants consistent with those known from QCD.  The nucleon is slightly more problematic, in the sense that with the Monte Carlo statistics accessible to us, the combined systematic and statistical errors are somewhat larger than in the pion case. We show that these data are completely consistent with recent lattice QCD calculations extending well into the chiral regime and having substantially smaller errors.   Fitting the lattice data with $\chi$pt again yields sensible low energy constants, so that although our IILM data are not sufficient to independently determine these constants, since our data is consistent with the lattice data, it is also consistent with these low energy constants.

Having carefully verified the consistency of the IILM results with lattice results and with the behavior expected from $\chi$pt, we conclude that the low energy effective theory indeed gives a useful approximation to the QCD path integral and hence use this model to explore the role of instanton induced dynamics.  An essential  quantity in the IILM is the overlap matrix $T_{ij}$ specifing the probability that a quark hops from instanton $i$ to instanton $j$, and we find that a value of the quark mass $m^*$ defined to be equal to an average value of $T_{ij}$  is a key parameter in characterizing the quark mass dependence of low energy QCD.  In the IILM, $m^*$ = 80 MeV, corresponding to $m_{\pi}$ = 600MeV, and characterizes the scale for three transitions.  It is the mass scale above which the fermion determinant is suppressed, the zero modes become subdominant, and the density of quasi-zero modes becomes independent of quark mass.  The significance of these transitions is discussed in the text.

The paper is organized as follows. In the next section, we review some of the basic aspects of the IILM. 
In section~\ref{dirac}, we present our analysis of the Dirac spectrum and compare our results with chiral perturbation theory predictions.  In section \ref{mass}, we present the details of our calculations of hadron masses and compare our results with lattice data. In section~\ref{extrapolation}, we perform chiral extrapolations of the nucleon and pion masses and compare the effective parameters of the low-energy chiral perturbation theory of the IILM with those of QCD.
In section~\ref{chiraldynamics}, we analyze how in the IILM the chiral dynamics is incorporated in the structure of the quark propagator and in the fermionic determinant, and we define and compute the mass scale $m^\star$ that sets the boundary of the chiral regime.  All results are summarized in section \ref{conclusions}.

\section{The Interacting Instanton Liquid Model}
\label{IILM}

In the IILM,  the QCD path-integral over the gauge field configurations is replaced by a sum over the configurations of a grand-canonical statistical ensemble of instantons and antiinstantons:
\begin{equation}
\label{eq:partf}
\mathcal{Z}_{QCD}\simeq\mathcal{Z}_{ILM} =\sum_{N_+, N_-} \frac{1}{N_+!N_-!}\int\prod_i^{N_++N_-}\textrm{d}\Omega_id(\rho_i)e^{-S_{int}}\prod_i^{N_f}\textrm{det}(i\sc{D}+im_f).
\end{equation}
Here, $\textrm{d}\Omega_i=\textrm{d}U_i\textrm{d}^4z_i\textrm{d}\rho_i$ is the measure in the space of collective coordinates, color orientation, position and size, associated with the single instantons. Quantum fluctuations are included in Gaussian approximation, through the semi-classical instanton amplitude $d(\rho_i)$.	 In what follows we will work in the thermodynamical limit and adopt the canonical formulation with $N_{+}=N_{-}=N/2$. 

Since instanton-instanton interactions are important to remove large-sized instantons from the vacuum~\cite{shuryak82}, the partition function (\ref{eq:partf}) describes an interacting liquid, rather than a dilute gas of pseudoparticles. 
The corresponding interaction action is defined as
\begin{eqnarray}
S_{int}\equiv S_{tot}-(N_+ + N_-) S_{0},
\end{eqnarray} 
where $S_{tot}$ is the total classical action corresponding to a given ensemble configuration and $S_{0}=\frac{8\,\pi}{g^2}$ is the individual single-instanton action in the dilute-gas limit.

In the IILM, the interaction action is approximated by a pure two-body instanton-antiinstanton term, which only depends on the relative coordinates of two pseudo-particles:
\begin{eqnarray}
S_{int}=\frac{1}{2}~\sum_{I\neq J}^{N/2} S_{int}(I,J)
\end{eqnarray}
Such a two-body action is calculated classically, by first determining the total action $S_{tot}$ from the  gauge potential $A_\mu(I,A)$ corresponding to the instanton-antiinstanton pair and then subtracting the free contribution $2 S_0$. Since no exact instanton-antiinstanton pair solution $A_{\mu}(I,A)$ to the classical Yang-Mills equations of motion is known, we rely on the stream-line construction, in which the classical action is minimized in all the directions except along the collective coordinate describing the separation between the two instantons \cite{stream}. 
The analytic instanton--antiinstanton gauge potential in the streamline {\it  Ansatz} is given by
\begin{equation}
	A_{\mu}^a= 2\eta_{\mu\nu}^a\frac{x_{\nu}}{x^2+\rho^2\lambda}+2R^{ab}\eta_{\mu\nu}^b\frac{\rho^2}{\lambda}\frac{1}{x^2(x^2+\rho^2/\lambda)}
\end{equation}	
where $\lambda$ is the streamline conformal parameter defined by 
\begin{eqnarray}
\lambda &=& \frac{R^2+\rho_I^2+\rho_A^2}{2\rho_I\rho_A}+\Big(\frac{(R^2+\rho_I^2+\rho_A^2)^2}{4\rho_I^2\rho_A^2}-1\Big)^{1/2}\\\nonumber
R &=& |z_I-z_A|
\end{eqnarray}
and 
\begin{equation}
	R^{ab}=\frac{1}{2}\textrm{Tr}[U^{\dagger}\tau^aU\tau^b]
\end{equation}
represents the relative color orientation of the two instantons.

A major shortcoming of the streamline construction is that, in the most attractive color orientation channel, the interaction action smoothly approaches $S_{int}=-2 S_0$ at short distances. This means that the instanton-antiinstanton pair tends to annihilate. 
The resulting gauge field configuration is characterized by a vanishing topological charge and a weak gauge field potential, and therefore it corresponds to a perturbative fluctuation.  Since perturbative contributions cannot be  treated consistently in the present approach, Shuryak and Sch\"afer~\cite{IILM}  suggested removing them by introducing a purely phenomenological short-range repulsive core:
\begin{equation}
\label{core}
	S_{core}=\frac{A}{\lambda^4}|u|^2,
\end{equation}
where $u$ is a color orientational factor. Such a term provides a cut-off to the momentum that can be exchanged through the instanton field, hence restricting the region of applicability of the approach to the non-perturbative sector characterized by momenta of the order 
$p< 1/\bar{\rho}$, where $\bar{\rho}$ is the average instanton size. In the language of effective field theory, this repulsive core plays the role of  a counterterm, parametrizing the ultraviolet physics. However, it should be stressed that our hard-core is not derived from a systematic contruction and therefore introduces model dependence in the calculation. The coefficient $A$ in (\ref{core}) controls the strength of the repulsion and is the only phenomenological parameter of the model. In our calculations we adopted the value $A=128$ suggested by Sch\"afer and Shuryak~\cite{IILM}. 

Once the interaction action is defined, it is possible to compute the main properties of the ensemble, the instanton density $n$, and  instanton size distribution $d(\rho)$, by minimizing the ensemble's free energy numerically. One can then calculate arbitrary Euclidean Green functions by performing   Monte Carlo averages over instanton configurations, using the Metropolis algorithm. 
As in analogous  lattice QCD calculations, one must set the scale in physical units by matching one dimensionful quantity.

In order to compute correlation functions involving quark field operators, we follow the same prescription adopted in lattice simulations, i.e. we first explicilty integrate-out the fermion fields and then compute Monte Carlo averages of the resulting Wick contractions. Such averages are performed using configurations which are obtained by an accept/reject Metropolis algorithm in which the contribution of the fermionic determinant is included in the Bolzmann weight,  corresponding to unquenched simulations.

Unlike in lattice QCD simulations, where the fermionic determinant and the quark propagators are evaluated using a purely numerical algorithm, in the IILM these quantities have a semi-analytic representation. This property allows us to identify the physical content of each term and establish a connection with the corresponding non-perturbative quark dynamics~\cite{rev}.
The fermionic determinant in a given instanton gauge field $A_\mu$ background is factorized into a contribution of near  zero-modes and a contribution arising from non-zero modes: 
\be
\textrm{Det}_A(\ds{D}+m_q)=\textrm{Det}_{A\,zm} \times \textrm{Det}^\prime_{A\,nzm}.
\label{det}
\ee 
The contribution of near-zero modes can be determined exactly, by expanding the Dirac operator on the basis of zero-modes of individual instantons:
\be
\textrm{Det}_{A\,zm}=\textrm{Det}(T+m_q),
\label{detzm}
\ee
where $T_{I J}= \int d^4 z \psi^{0\,\dagger}_I(z) i D_\mu \gamma_\mu \psi_j^0(z)$ is the overlap matrix which represents the probability amplitude for quarks to "hop" from the instanton $I$ to the instanton $J$.
The non-zero mode part of the fermionic determinant is approximated with the product of the non-zero mode contributions of each individual instanton: 
\be
\textrm{Det}^\prime_{A\,zm}=\prod_{i}^{N_++N_-}(1.34 m_q \rho_i)
\label{detnzm}
\ee

Simlarly, the quark propagator  consists of a zero-mode part and a non-zero mode part:
\begin{eqnarray}
S(x,y)_A=S_{zm}(x,y)_{A}+S_{nzm}(x,y)_{A}.
\end{eqnarray}
The zero-mode part of the propagator $S_{zm}(x,y)_A$  dominates in the low-energy regime and encodes the information on the physics of chiral symmety breaking and axial anomaly saturation. It is constructed by representing the Dirac operator in the basis of zero-mode wave functions $\psi^{0}(x)$ of the individual instantons and reads: 
\begin{eqnarray}
\label{Szm}
S_{zm}(x,y)_A=\sum_{I J} \psi_I^{0}(x)\psi_J^{0\dagger}(y)~\left[\frac{1}{\hat{T}-i m_q}\right]_{I J}.
\end{eqnarray}
 The non-zero-mode part is approximated as the sum of the non-zero-mode propagators in the field of the invidual instantons~\cite{rev}.


\section{The Spectrum of the Dirac operator and the chiral regime in the IILM}
\label{dirac}

 In the previous section, we have seen that the main approximation of the IILM consists in replacing the QCD path 
integral with a statistical sum over a classical ensemble of pseudoparticles. 
In this section, we address the question of whether such an approximation provides a realistic description of the non-perturbative quark-gluon dynamics  associated with the spontaneous breaking of chiral symmetry in QCD.
To this end we focus on the low-virtuality sector of the spectrum  of the Dirac Operator $\ds{D}$, which encodes information about the dynamics associated with the spontaneous breaking of chiral symmetry. Such a connection is made explicit in the Banks-Casher relation which relates the density of eigenvalues with small virtuality to the quark condensate:
\begin{equation}
\label{eq:bcr}
\langle \overline{q}q\rangle=-\pi\rho(\lambda=0),
\end{equation}
We recall that this relation holds in a infinite-volume system and in the chiral limit.
Our goal is to explore finite-mass corrections to the Banks-Casher relation in the instanton model and check if such corrections are consistent with predictions derived from $\chi$pt. 

In order to set the framework of our analysis, it is useful to briefly review how the Banks-Casher  relation (\ref{eq:bcr}) was obtained. The quark propagator $S(x,y)_A$ in the fixed background $A_\mu$ is expanded in terms of the eigenvalues $\lambda$ and eigenvectors $\psi_{\lambda}$ of the Dirac operator as
\begin{equation}
	S(x,y)_A=\sum_n\frac{\psi_n(x)\psi^{\dagger}_n(y)}{m_q-i\lambda_n}.
\end{equation}
Chiral symmetry implies that for every non-zero eigenvalue $\lambda$ with eigenvector $\psi_{\lambda}$ there is another eigenvalue $-\lambda$ with eigenvector $\gamma_5\psi_{\lambda}$. Hence, setting $x=y$ and using the orthonormality of the basis of eigenfunctions of $\psi_{\lambda}$ we obtain
\begin{equation}\label{eq:lim}
	\frac{1}{V}\int \textrm{d}^4x~Tr~[S(x,x)]_A=\frac{2m_q}{V}\sum_{\lambda_n>0}\frac{1}{m_q^2+\lambda_n^2}
\end{equation}
Introducing the spectral density $\rho(\lambda)=\langle\sum_{n}\delta(\lambda-\lambda_n)\rangle$ we can rewrite  (\ref{eq:lim}) as
\begin{equation}\label{eq:lim2}
	\langle\overline{q}q\rangle=-2m_q\int_0^{\infty}\textrm{d}\lambda\frac{\rho(\lambda)}{m_q^2+\lambda^2}
\end{equation}
At this point we take the limit $V\rightarrow\infty$ and $m\rightarrow 0$. We notice that the order in which the two limits are taken is crucial since  we only  have a finite quark condensate in the thermodynamic limit: 
\begin{equation}
	\langle\overline{q}q\rangle= \lim_{m_q\rightarrow 0}\lim_{V\rightarrow\infty}-2m_q\int_0^{\infty}\textrm{d}\lambda\frac{\rho(\lambda)}{m_q^2+\lambda^2} =-\pi\rho(\lambda=0) .
\end{equation}

Eq. (\ref{eq:bcr}) can be interpreted as the lowest-order term in the Taylor expansion of $\rho(\lambda)$ near the origin, in the chiral limit.
The next order in $\lambda$ was derived by Smilga and Stern~\cite{smilga} using $\chi$pt:
\begin{equation}\label{eq:smilga1}
	\rho(\lambda)=-\frac{1}{\pi}Bf_0^2+\frac{B^2(N_f^2-4)}{32\pi^2N_ff_0^2}|\lambda|+ \mathcal{O}(\lambda) , \\
\end{equation}
where $N_f$ is the number of flavors,  $B$ and $f_0$ are the constants that appear in the lowest order $\chi$pt Lagrangian
\begin{equation}\nonumber
	\mathcal{L}=f_0^2(\frac{1}{4}~Tr[\partial_{\mu}U\partial^{\mu}U^{\dagger}]+2B~Tr[MU^{\dagger}+UM^{\dagger}]) ,
\end{equation}
$U=\exp\big(i\frac{\phi(x)}{f_0}\big)$, and $M$ is the diagonal quark mass matrix.

\begin{figure}
\includegraphics[width=.82\textwidth]{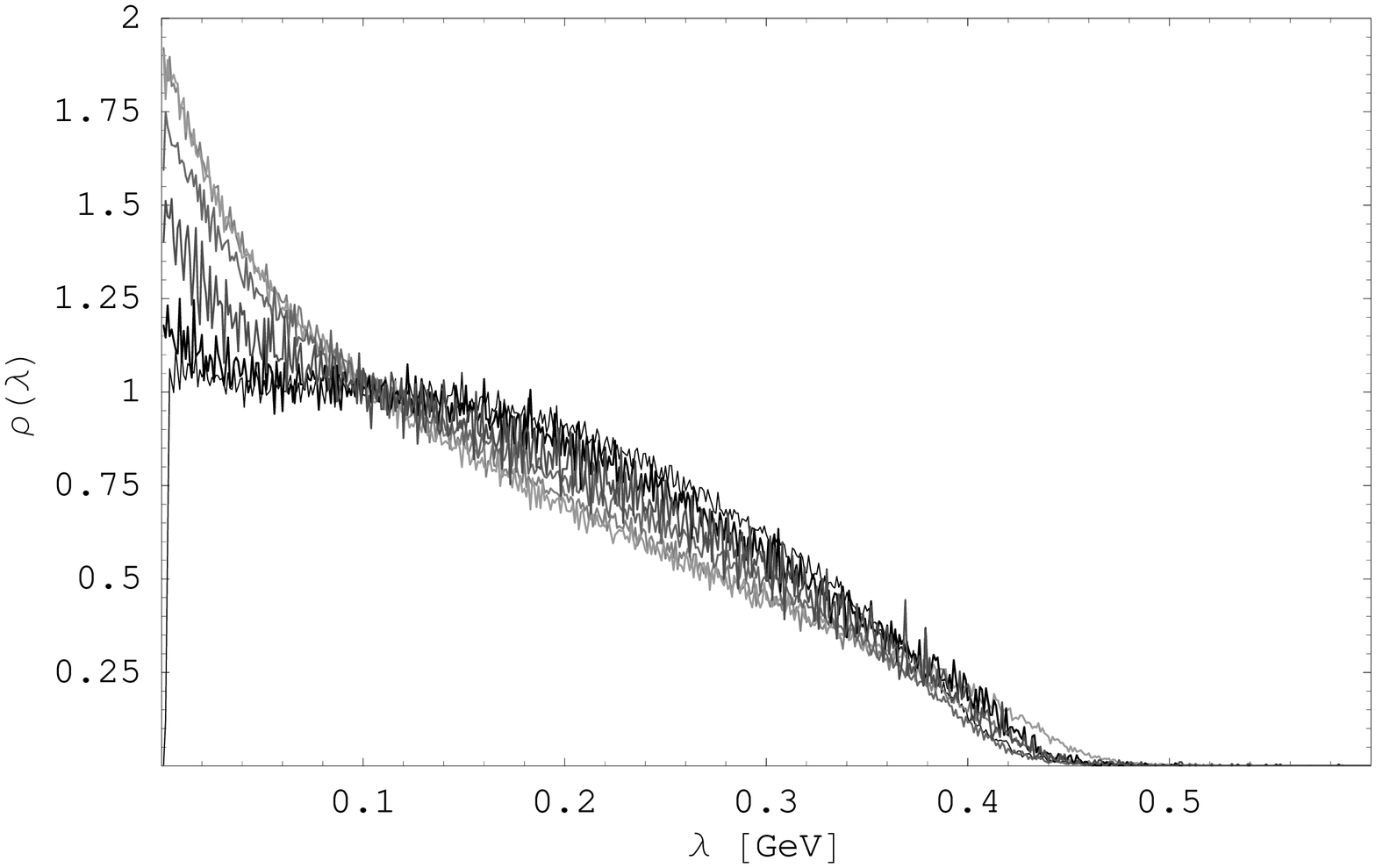}
\includegraphics[width=.88\textwidth]{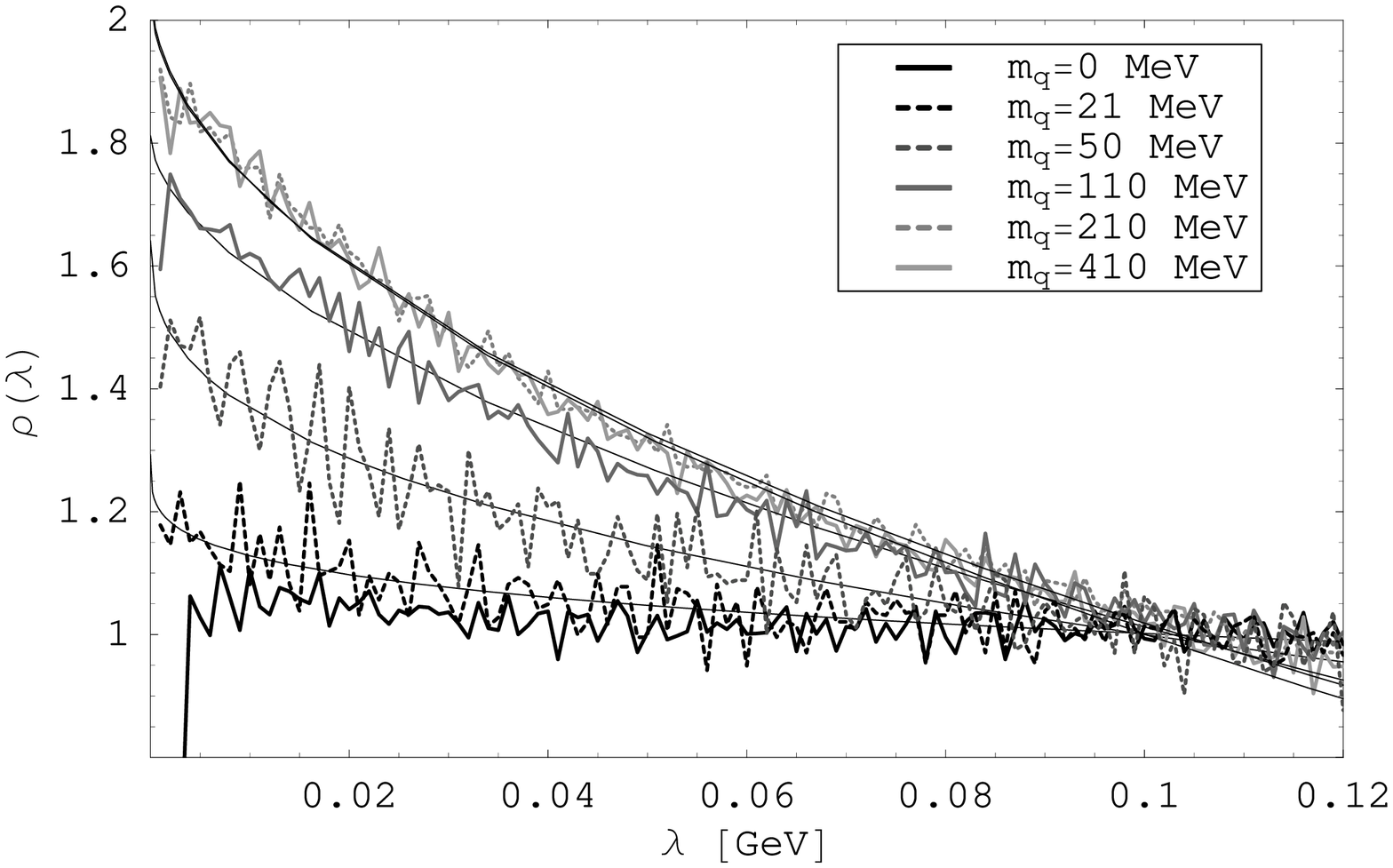}
	\caption{ Spectral density of the Dirac operator for $N_c=3$ and $N_f=2$.  The lower panel shows an enlargement of the upper panel in the region $0\leq\lambda\leq0.3$ and the solid lines represent best fits using $\rho(\lambda)= c_1+c_2\times  m_q^{\alpha}\lambda^{\alpha-1}$.}
	\label{fig:rla}
\end{figure}

We stress the fact that the result (\ref{eq:smilga1}) is valid only in the limit in which the quark mass is much smaller than the typical eigenvalue saturating the spectral integral (\ref{eq:lim2}),  $ m_q\ll\lambda$. 
If the quark mass is of the order of the typical eigenvalue $\lambda$, then mass-dependent corrections to the spectral density profile are expected to appear. Hence, the question arises if there exists  a range of masses for which the $\chi$pt predictions are accurate, yet mass corrections to (\ref{eq:smilga1}) become important.
In order to address this question, Smilga and Stern considered the 3-point scalar correlator: 
\begin{equation}
	K^{abc}=\int d^4x d^4y d^4z \langle 0|S^a(x)S^b(y)S^c(z)|0\rangle,
\end{equation}
where $S^a(x)$ is the scalar density operator, $S^a(x)=\bar{q}(x)\tau^a q(x)$.
By proceeding in the same way as for the quark condensate they obtained an expression for $K^{abc}$ in terms of the density of eigenvalues, $\rho(\lambda)$:
\begin{equation}\label{eq:smilga2}
	K^{abc}=-d^{abc}m_q\int_0^{\infty}\rho(\lambda)\frac{(m_q^2-3\lambda^2)}{(m_q^2+\lambda^2)^3}\textrm{d}\lambda.
\end{equation}
In the limit in which all mass corrections to (\ref{eq:smilga1}) can be neglected, one can obtain a prediction for the 3-point scalar correlator by substituting  the expansion of $\rho(\lambda)$ given in (\ref{eq:smilga1}) in (\ref{eq:smilga2}) and performing the integral. The constant term in $\rho(\lambda)$ does not contribute, while the linear term gives rise to a power-law singularity in the pion mass:
\begin{equation}
K^{abc}= \frac{B^3(N_f^2-12)}{64\pi^2m_{\pi}^2N_f}\frac{1}{m_q}\propto 1/(m_{\pi}^2) .
\label{smilgaK}
\end{equation}
On the other hand, evaluating the same correlator to lowest-order in $\chi$pt one finds: 
\begin{equation}
	K^{abc}=\frac{B^3(N_f^2-12)}{64\pi^2m_{\pi}^2N_f}d^{abc}.
\end{equation}

Note that, although in both cases one finds the same power-law dependence on the pion mass, the numerical coefficient turns out to  be different
\begin{equation}
	N_f^2-4\neq(N_f^2-12)/2
\end{equation}
The explanation of the mismatch is that the characteristic eigenvalues, $\lambda$, saturating the power divergent integral (\ref{eq:smilga2}) are of order $m_q$, so we are out of the range of validity of (\ref{eq:smilga1}). 
In order to reproduce the correct $\chi$pt prediction for the scalar 3-point correlator such mass corrections have to be included. 

It is particularly interesting to consider the case  $N_f=2$, for which the linear term in (\ref{eq:smilga1}) vanishes and   the distribution of eigenvalues near the origin is flat. In this case, the power-law divergence in the pion mass predicted by $\chi$pt must arise entirely from mass corrections. 
This observation can be used to check the consistency of the IILM with $\chi$pt. We have performed simulations of the Dirac spectrum at different values of the quark mass, ranging from $0$ to $400$ MeV.
The contribution of the near zero-mode zone to the $N_f=2$ spectral density of the positive eigenvalues is presented in Fig.\ref{fig:rla}. 

Some comments on these results are in order. When the quark mass approaches zero, our results become consistent with the flat trend predicted by (\ref{eq:smilga1}):
\be
\lim_{m_q\to0}\lim_{\lambda \to0} \rho(\lambda) =  \textrm{Const} + O(\lambda^2).
\ee  
The value of the constant  depends on the normalization of the spectrum and can be fixed by imposing the condition (\ref{eq:bcr}).
This case has been already considered in previous studies of Shuryak and Verbaarschot \cite{ds1,ds2}. On the other hand, as the quark mass increases, the structure of the spectrum changes appreciably. The flatness near $\lambda=0$ disappears and a peak near the origin develops. In Fig.\ref{fig:rla}  we see that for small $\lambda$, all curves at different masses are very well fitted with functions of the form $\rho(\lambda)-\rho(0)\propto m^{\alpha}\lambda^{1-\alpha}$. If this form is introduced in the spectral representation of the scalar three-point function (\ref{smilgaK}), the integral can be carried-out analitically leading to results proportional to the inverse quark mass.

To directly compare the $\chi$pt formula for $K^{abc}$ with the results of IILM calculations, it is convenient to consider the combination $m_q\frac{K^{abc}}{\rho_{m=0}(0)}$. In fact, using the Gell-Mann Oaks Renner and Banks-Casher relations, we find the simple $\chi$pt prediction  
\be\label{eq:b0ovf0}
	m_q\lim_{m_q\rightarrow0}\frac{K^{abc}}{\rho_{m_q=0}(0)}=\frac{B(N_f^2-12)}{128\pi f_0^2N_f}=\textrm{const}.
\ee

On the other hand, from the spectral representation (\ref{eq:smilga2}) we find 
\be\label{eq:bf0}\nonumber
	m_q \frac{K^{abc}}{\rho_{m_q=0}(0)}&=& -m_q^2\int\textrm{d}\lambda\frac{\rho_{m_q}(\lambda)}{\rho_{m_q=0}(0)}\frac{(m_q^2-3\lambda^2)}{(m_q^2+\lambda^2)^3}d^{abc} ,
\ee
which can be computed directly from our IILM points. 
\begin{figure}
\includegraphics[width=1.\textwidth]{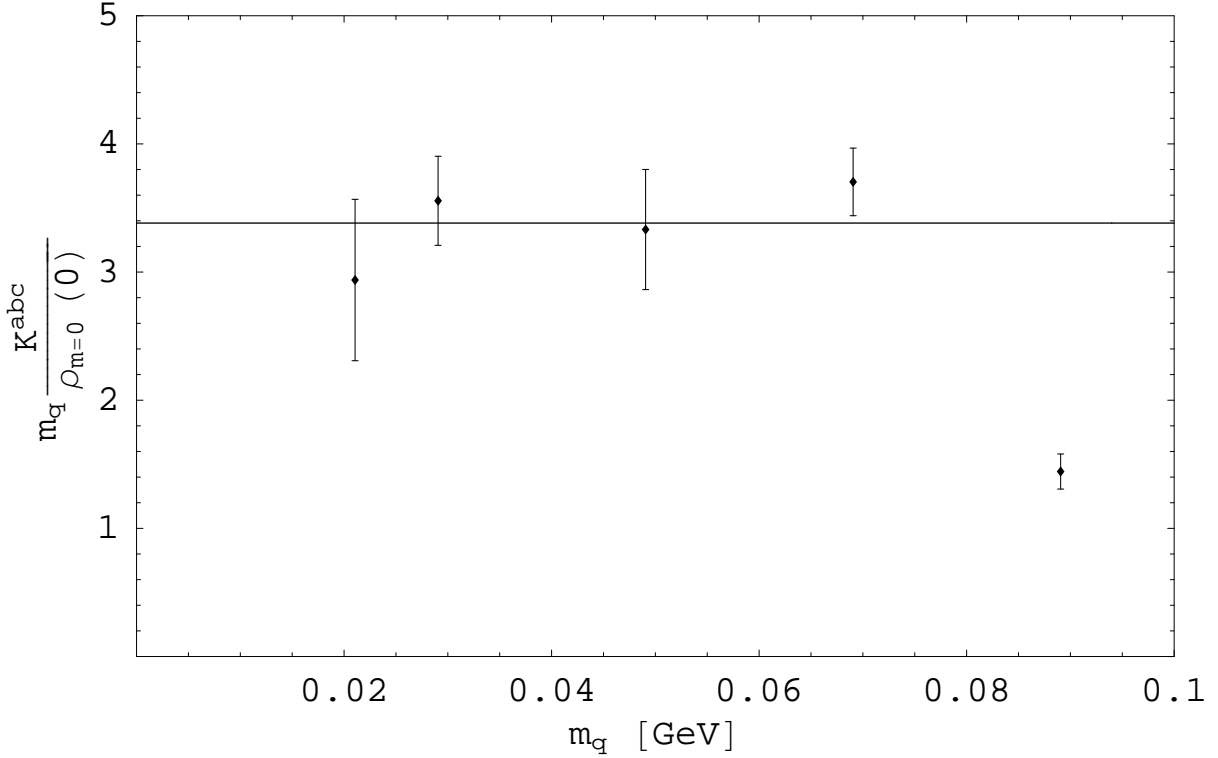}
	\caption{$m_q\frac{K^{abc}}{\rho_{m=0}(0)}$ at different quark mass values in the IILM.}
	\label{fig:bf0}
\end{figure}

The results are plotted in Fig.\ref{fig:bf0}, and we observe that the expected constant behavior with $m$ is obtained for $m_q<80$~MeV. Thus, in this case,  the IILM  reproduces   the structure of  the Dirac spectrum dictated by chiral symmetry.

\section{Pion and Nucleon Mass in the IILM}\label{mass}

In this section we present our IILM calculation of the nucleon and pion masses performed at different values of the quark masses and we compare with the available lattice results. 
Exploratory estimates of the nucleon and pion masses in the IILM were performed at a relatively large quark mass in \cite{IILM2}. However, such calculations were based on the analysis of short-range point-to-point Euclidean correlators and relied on a specific assumption for the spectral density (i.e. pole-plus-continuum parametrization).
In the present work, we have chosen to avoid any {\it a priori} assumption on the spectral density. Instead, we have performed the  effective-mass plot analysis, which is routinely used  to extract the lowest-lying hadron masses from lattice QCD simulations.

To compute  the mass of the pion, we have evaluated the   correlation function: 
\begin{equation}
G_{\pi}(\tau)=\int\textrm{d}^3{\bf x}\langle 0|T[j_5({\bf x},\tau)\overline{j}_5({\bf 0},0)|0\rangle, 
\label{Gpion}
\end{equation}
where $j_5(x)$ is the pseudoscalar density operator
\begin{equation}
j_{5}^a(x)=\overline{q}(x)\gamma_5\tau^a q(x).
\end{equation}
To compute the nucleon mass we have evaluated the correlation function 
\begin{equation}
G_{N}(\tau)=\int\textrm{d}^3{\bf x}\langle 0|T[j_N({\bf x},\tau)\overline{j}_N({\bf 0},0)P_+|0\rangle,
\label{GN}
\end{equation}
where $P_+=\frac{1-\gamma_4}{2}$ is the positive-parity projector and
\begin{equation}
j_{N}^a(x)=\varepsilon^{abc}~u^a(x) C\gamma_5 d^b(x) u^c(x).
\end{equation}

The integration over the final position ${\bf x}$  in (\ref{Gpion}) and (\ref{GN}) ensures projection onto zero-momentum states. 
The mass of these hadrons can then be extracted from the plateau in  the large Euclidean time limit of the effective mass, i.e. using 
\be
\label{eq:efmp}
M_{\pi/N} &=& \lim_{\tau\rightarrow\infty} M^{eff}_{\pi/N}(\tau)\\
M^{eff}_{\pi/N}(\tau) &=& \frac{1}{\Delta\tau}\ln\frac{G_{\pi/N}(\tau)}{G_{\pi/N}(\tau+\Delta\tau)}.
\ee

The point-to-point correlation functions in (\ref{Gpion}) and (\ref{GN}) were computed by performing Monte Carlo averages over instanton ensemble configurations, as described in section {\ref{IILM}.  The momentum-projection integrals in (\ref{Gpion}) was carried-out using an adaptive Monte Carlo routine (VEGAS) with two iterations of 6000 points. For the momentum-projection of the nucleon 2-point function (\ref{GN}) we combined the results of VEGAS with those obtained using a cubic grid with grid spacing $0.1$ fm.
Statistical errors over ensemble averages were obtained using the jackknife technique, with a bin size of $10$ configurations. 
We have used five sets of 250  independent configurations corresponding to quark masses ranging from 20 to 90 MeV.  
In order to reduce finite volume artifacts, we performed our computations in two boxes of size $3.45^3\times 5.9\ \textrm{fm}^4$ ---for the two lightest quark masses--- and $2.96^3\times 5.9\ \textrm{fm}^4$ ---for the other masses---. We have checked that with such a choice the condition $m_{\pi}L>5$ was always satisfied.

All simulations where performed using the 1TFlop/s ECT* cluster, for a total computational time of $30000$ cpu hours. In Fig.~\ref{fig:pn} we present  pion and nucleon effective mass plots for the different values of quark mass. 
 We observe that the all pion effective mass plots display a very clean plateau, from which it is possible to unambiguously  read-off the pion mass from a correlated chi-square fit. On the other hand, the interpolation of the effective mass plots of the nucleon is somewhat more problematic. In fact, not only statistical errors are larger, but also some of the plots are consistent with  a slight slope, in the large Euclidean time regime. Hence,  one necessarily needs to account for the systematic errors, which can be estimated from the smallest and largest mass compatible with the effective mass plot, in the nearly flat region $\tau\gtrsim~1$~fm.
\begin{figure}
		\centering
		\subfigure{\includegraphics[width=.38\textwidth]{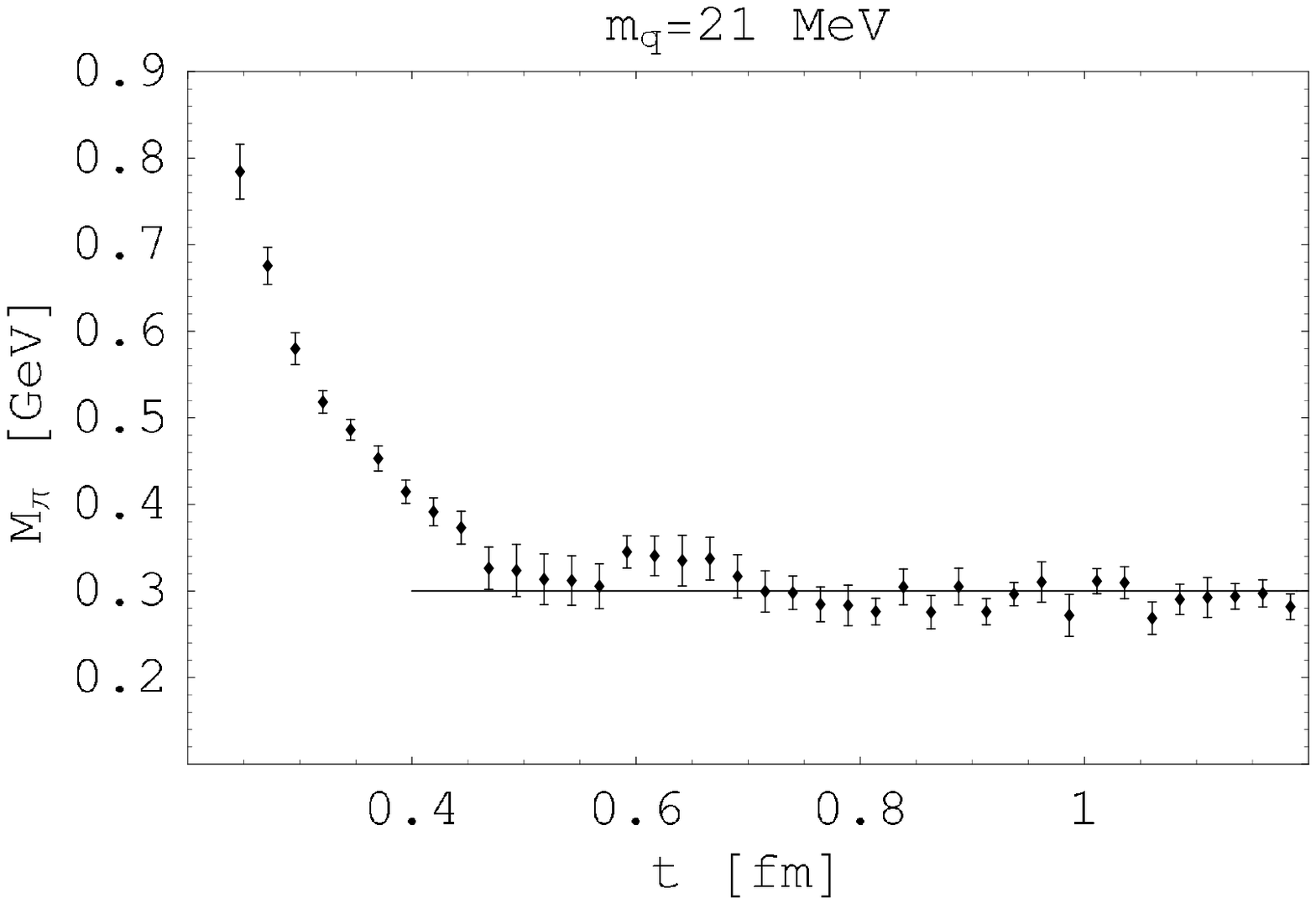}}\hspace{5mm}%
		\subfigure{\includegraphics[width=.38\textwidth]{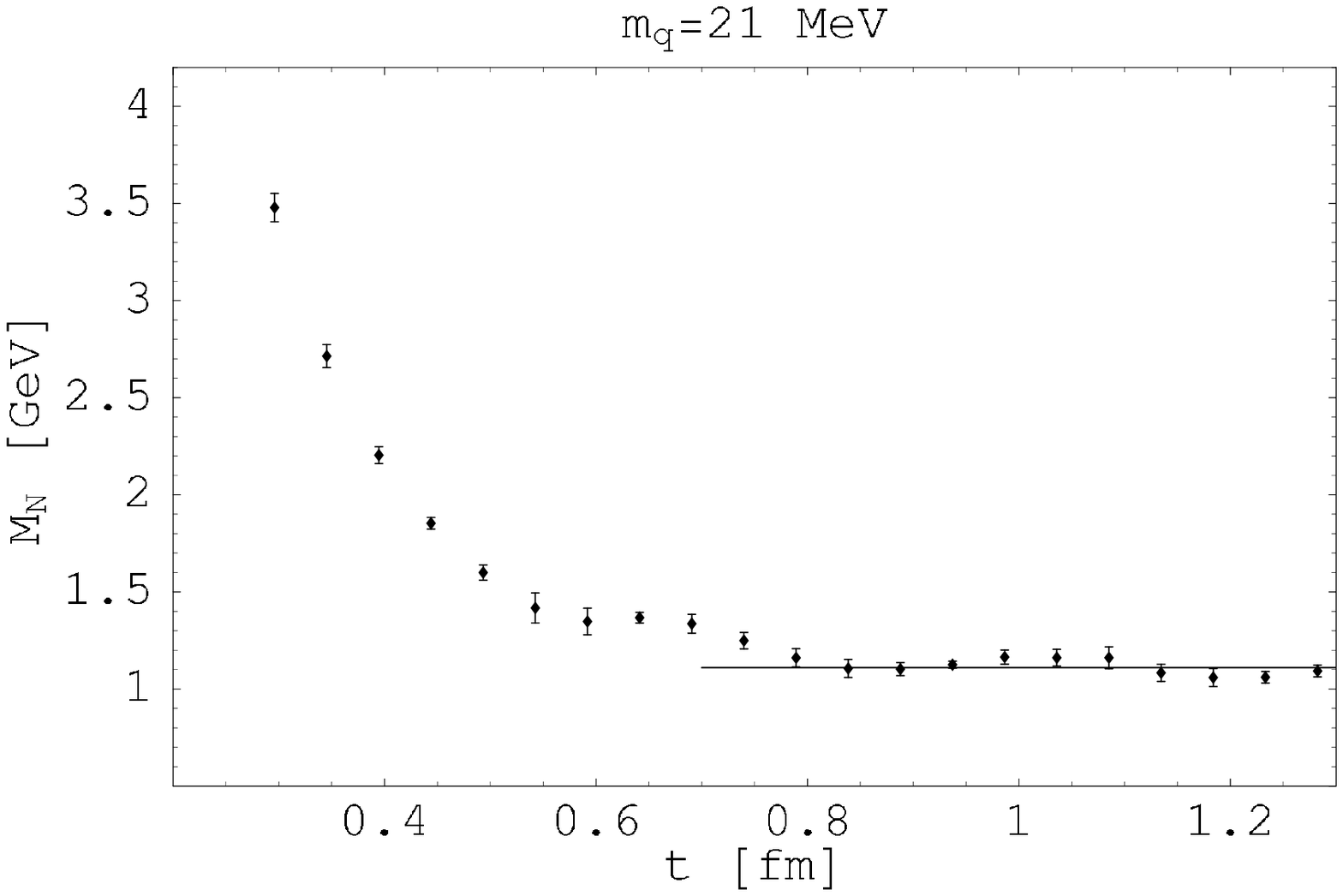}}
		\subfigure{\includegraphics[width=0.38\textwidth]{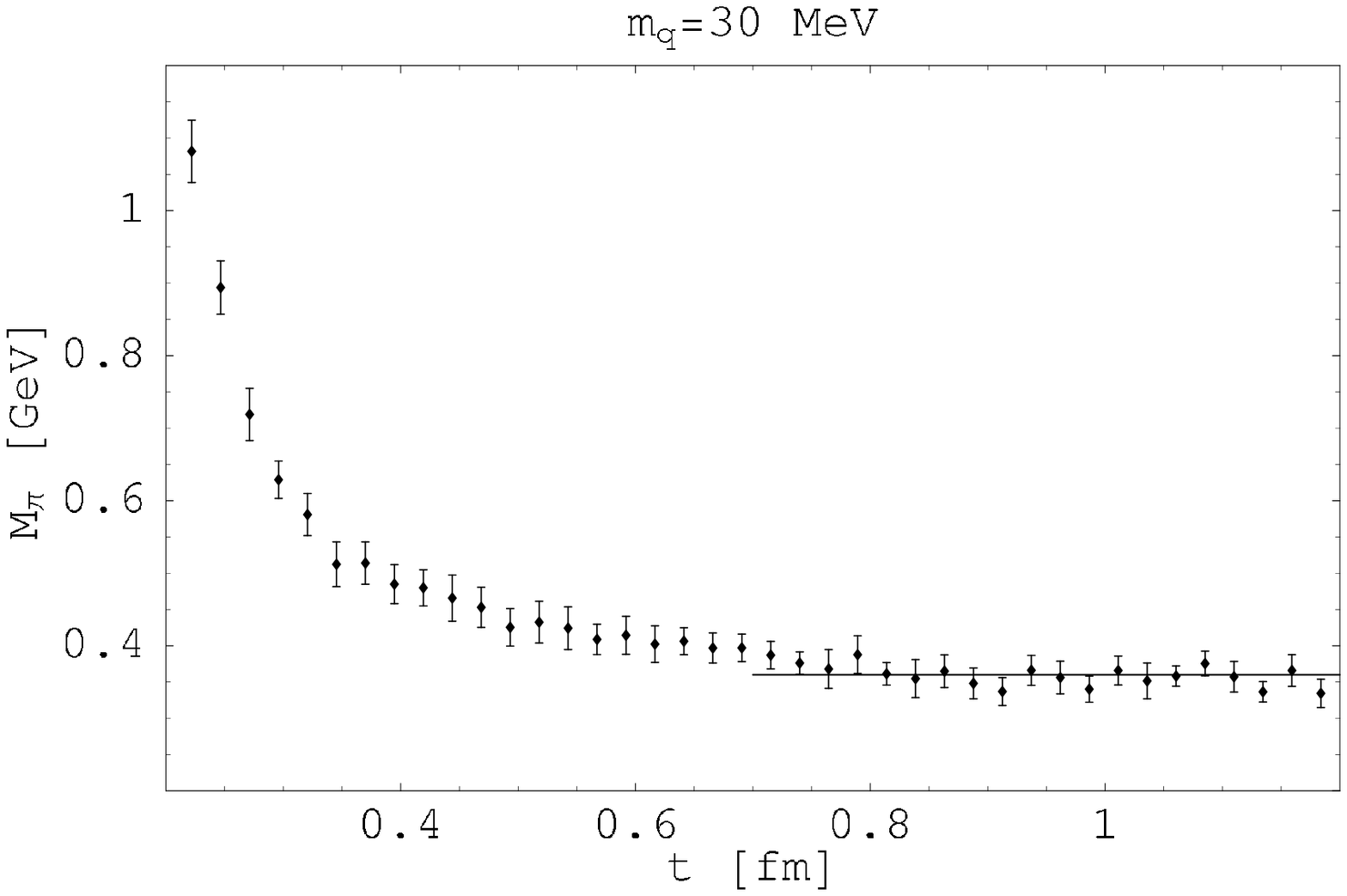}}\hspace{5mm}%
		\subfigure{\includegraphics[width=0.38\textwidth]{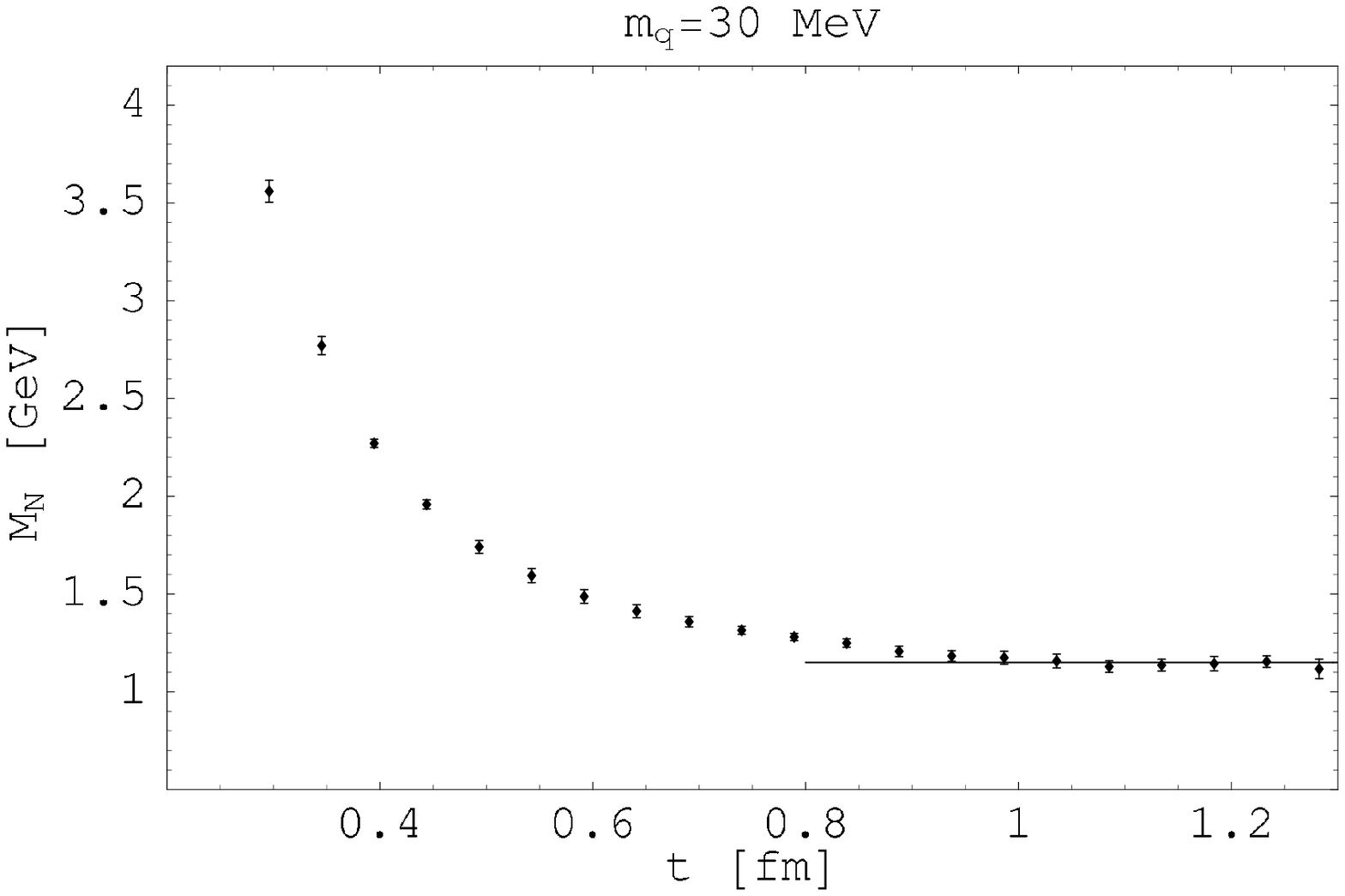}}
		\subfigure{\includegraphics[width=0.38\textwidth]{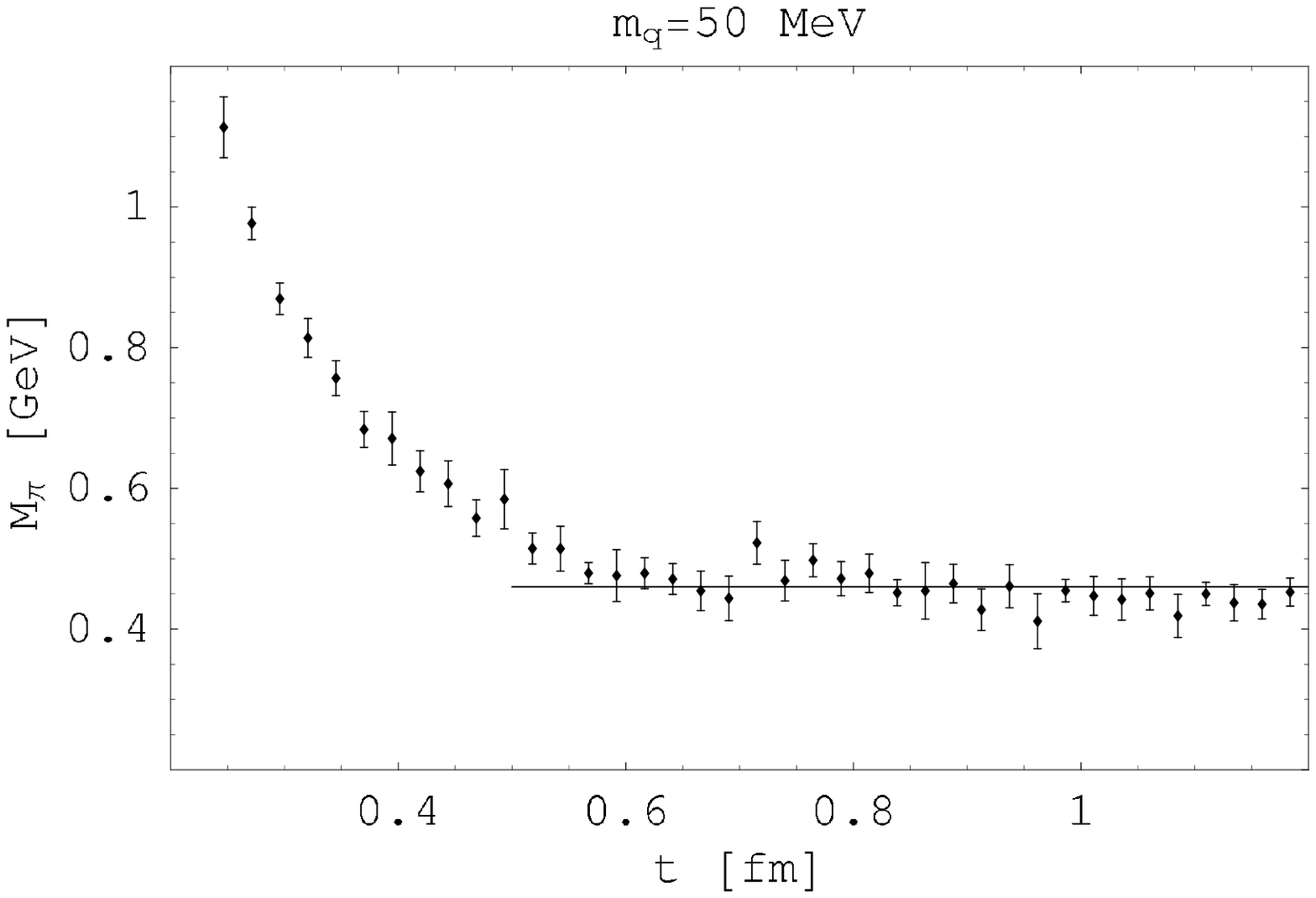}}\hspace{5mm}%
		\subfigure{\includegraphics[width=0.38\textwidth]{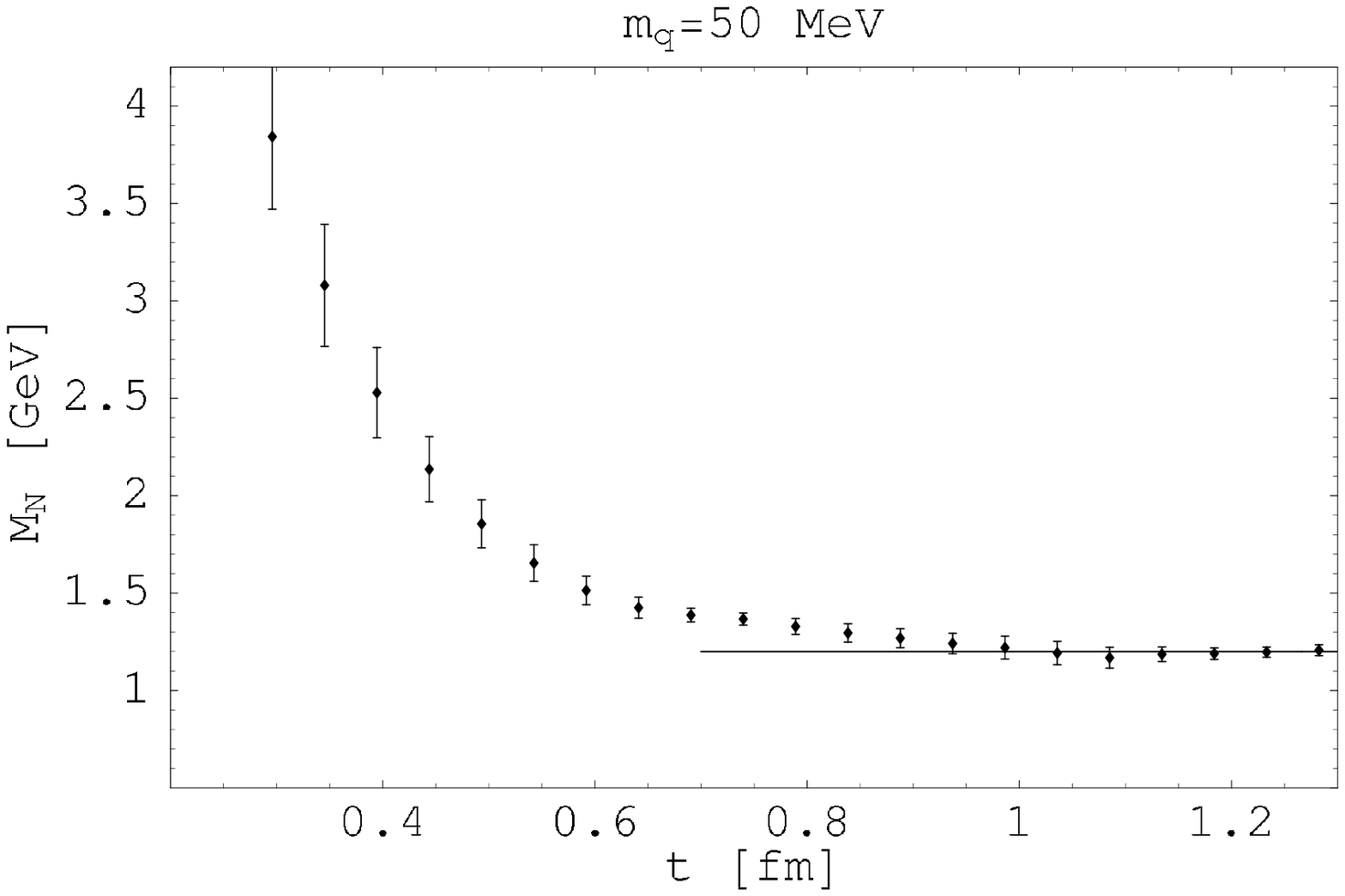}}
		\subfigure{\includegraphics[width=0.38\textwidth]{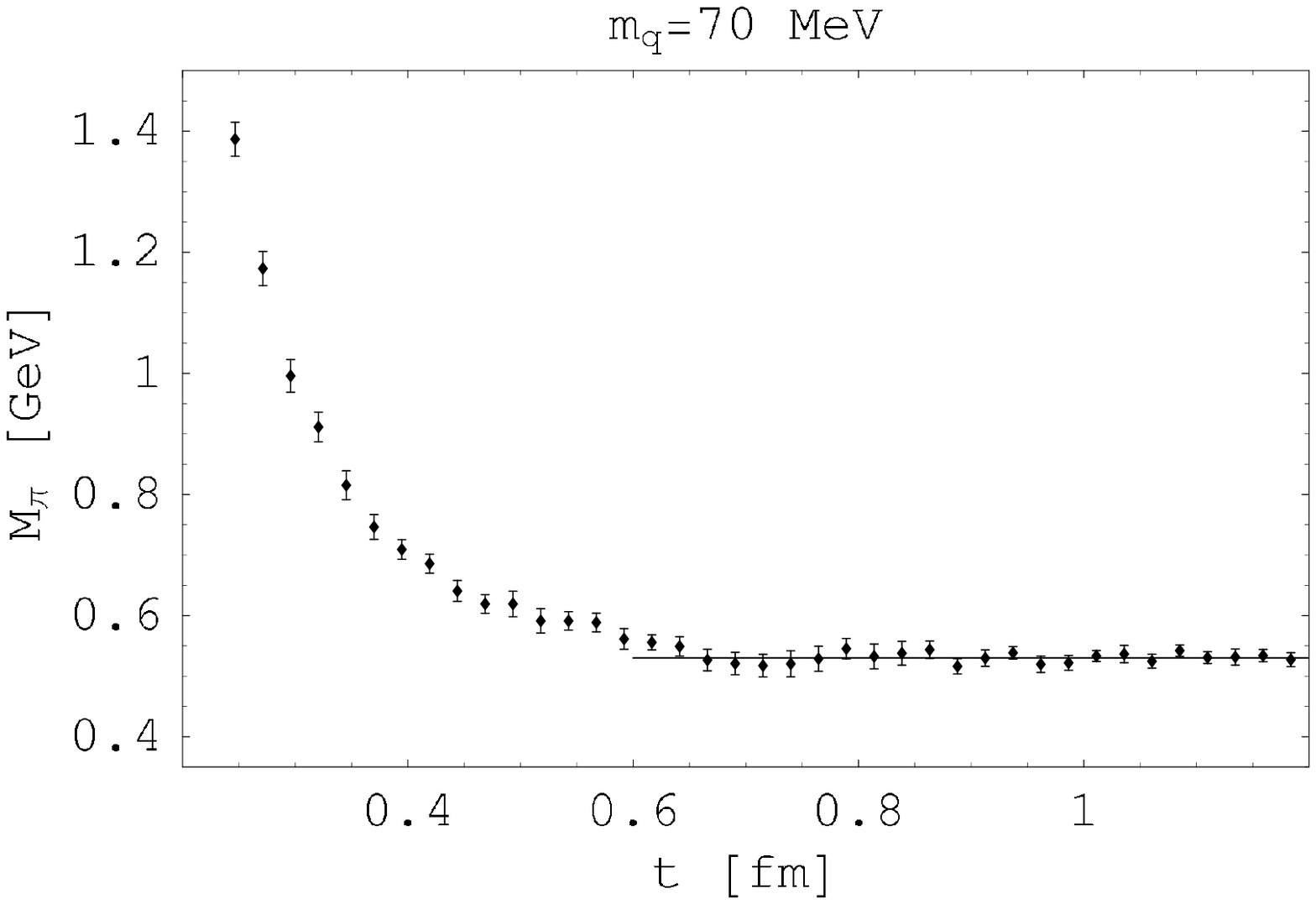}}\hspace{5mm}%
		\subfigure{\includegraphics[width=0.38\textwidth]{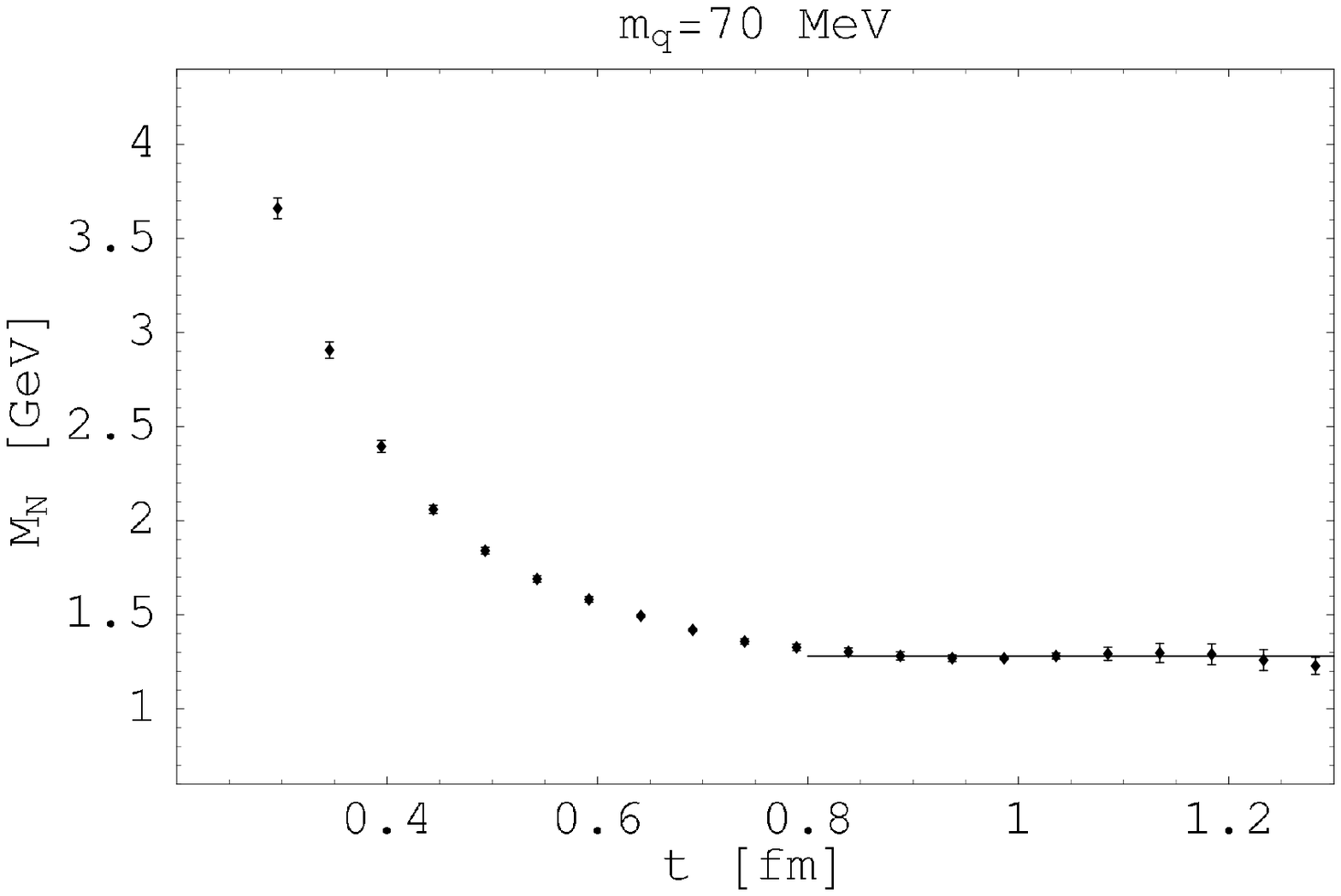}}
		\subfigure{\includegraphics[width=0.38\textwidth]{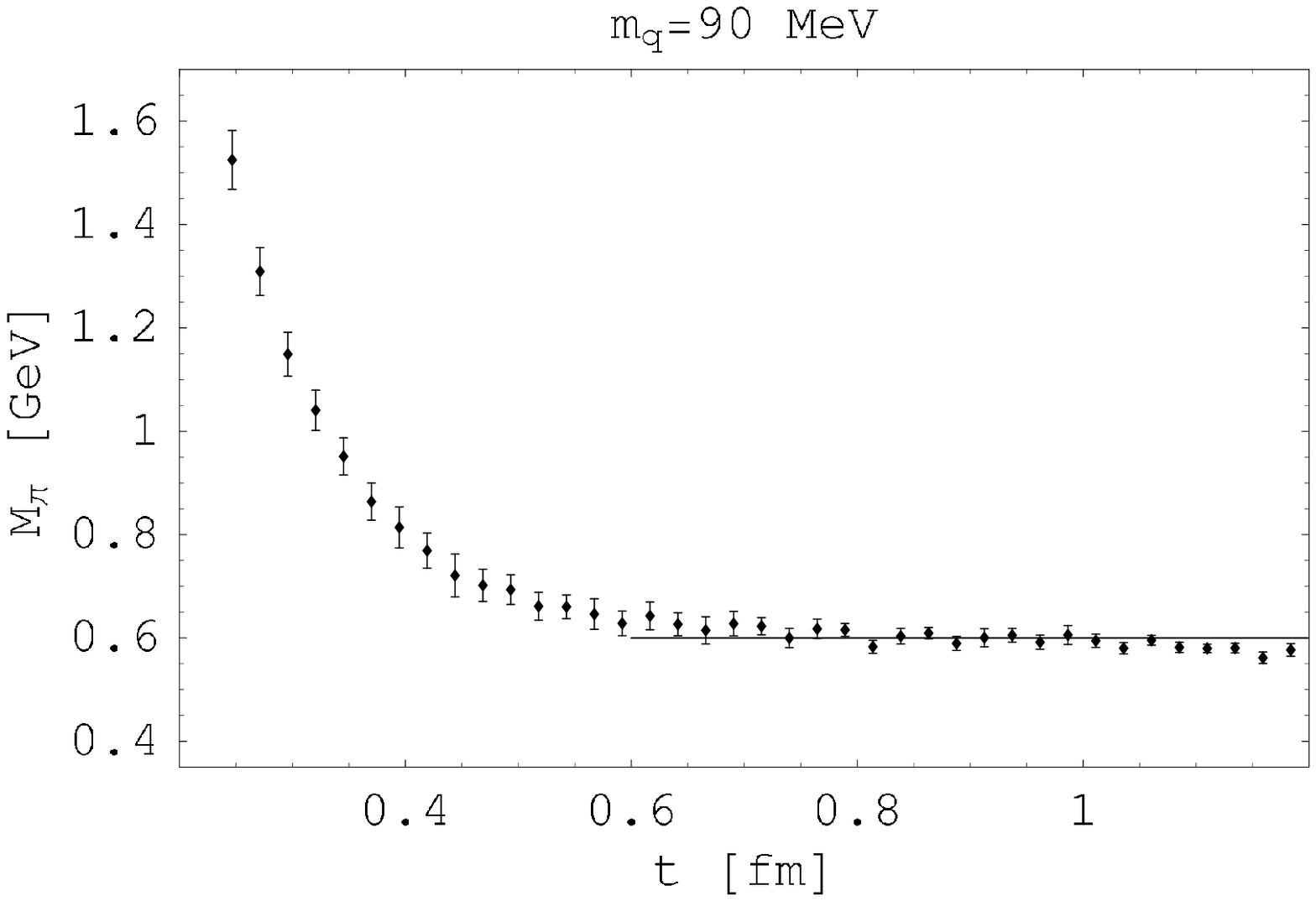}}\hspace{5mm}%
		\subfigure{\includegraphics[width=0.38\textwidth]{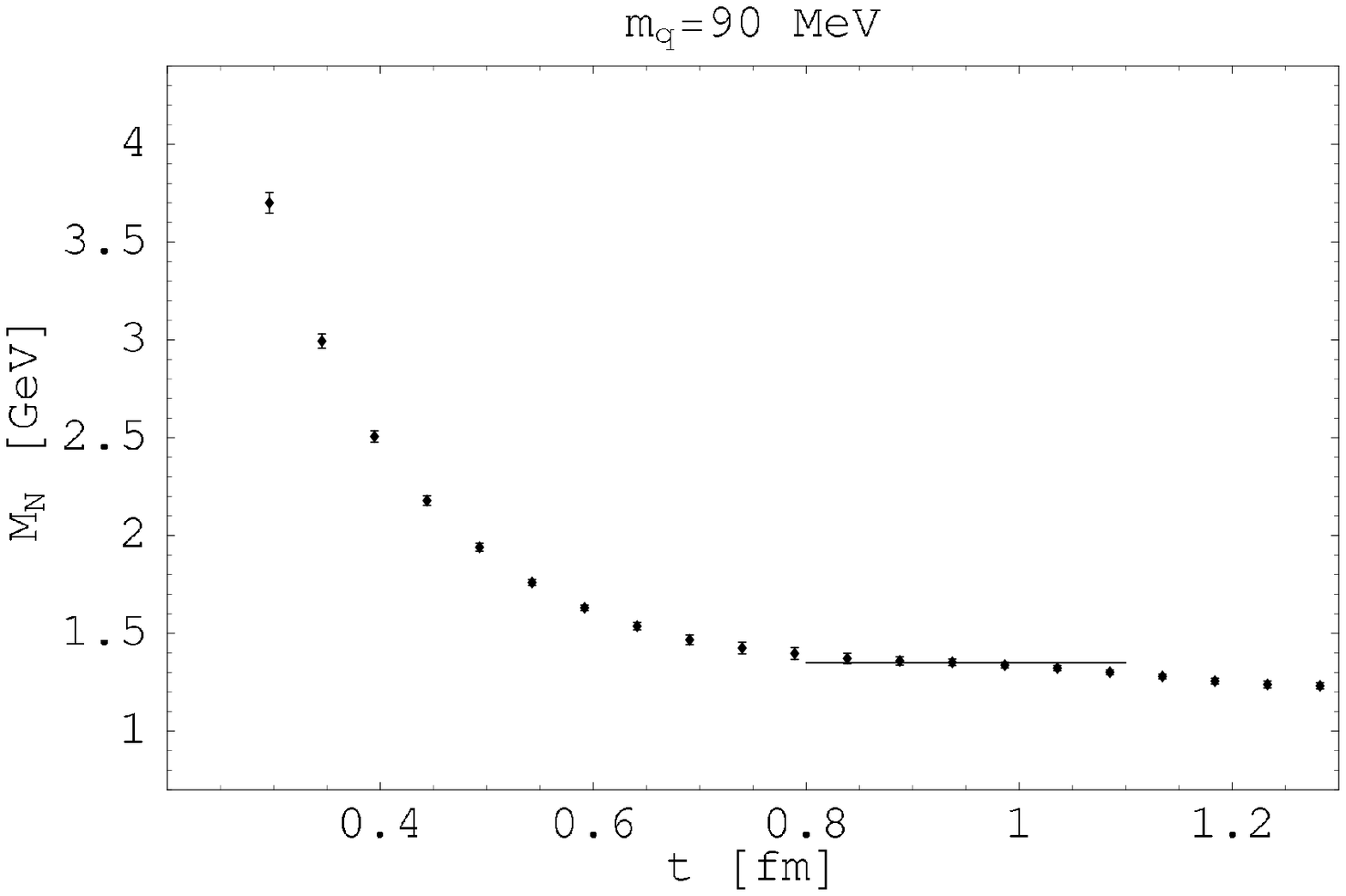}}
	  \caption{ Pion and nucleon effective mass plots in the IILM at different quark masses}\label{fig:pn}
\end{figure}

We recall that in the instanton model, the average instanton size $\bar{\rho}$ is the parameter determining the ultraviolet cut-off and playing the role of the lattice spacing $a$ in lattice gauge theory. 
Hence,  we have expressed all dimensional quantities in units of 
 $\bar{\rho}$ and then determined it from a matching condition.   For simplicity, we have chosen the condition that the IILM prediction for the nucleon mass  
should match the CP-PACS lattice result at the pion mass $m_\pi$~=~525~MeV. The values of the nucleon masses for $m_\pi<500$~MeV are therefore predictions of the model.
The choice of the matching point was motivated by the fact that  the pion mass $\simeq~500$~MeV  represents an upper bound for the mass regime of QCD where the dynamics of chiral symmetry breaking is expected to play an important role.

In Fig. \ref{fig:massr} we present the results of our calculations for the nucleon mass as a function of the pion mass squared for different choices of the cut-off $\bar{\rho}$. 
\begin{figure}
	\centering
		\includegraphics[width=1.\textwidth]{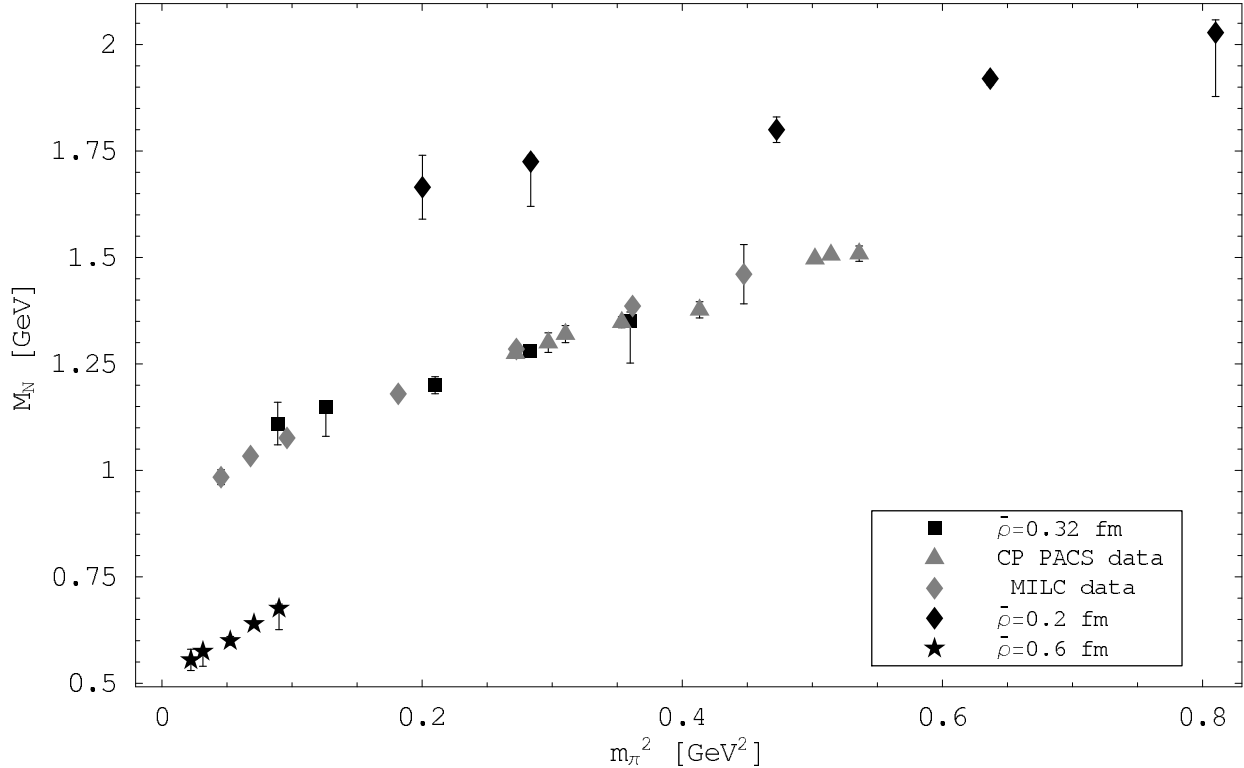}
	\caption{  Comparison between nucleon masses obtained with the IILM for three different choices of the average instanton size $\bar{\rho}$ and those calculated on the Lattice by the  CP-PACS~\cite{latticed} and MILC~\cite{Aubin:2004wf} collaborations.}\label{fig:massr} 
\end{figure}
Comparing these IILM data at different $\bar{\rho}$ with the lattice data collaborations~\cite{latticed,Aubin:2004wf} clearly shows that agreement between IILM and lattice QCD predictions is obtained  only for a unique value of the cut-off scale, corresponding to an average instanton size of $0.32$ fm,  in excellent agreement with the early phenomenological estimate $\bar{\rho}\simeq 1/3$~fm~\cite{shuryak82} and recent lattice results~\cite{latticerho1,latticerho2,latticerho3,latticeILM1,latticerho4}.
The complete set of values of masses extracted from our IILM calculations is presented in Table~\ref{table:hadr}.
\vspace{1cm}
\begin{table}[h!]
\caption{Pion and nucleon masses fitted by effective mass plot with $\chi^2/ndf\leq1$. The quark masses are determined at a scale $2\ \GeV$ as in \cite{musakh,bowman}}
\begin{center}\vspace{.2cm}
\begin{tabular}{lll}\hline\hline
	\rule{0pt}{3ex} $m_q$ [GeV]& Pion [GeV]&Nucleon [GeV]\\ \hline
  \rule{0pt}{3ex}
	$0.021 $&$ 0.300\pm 0.004 $&$ 1.11^{+0.05}_{-0.05} $\\
	$0.03 $&$ 0.360\pm 0.004 $&$ 1.15_{-0.07}^{+0.01} $\\
	$0.05  $&$ 0.460\pm 0.004 $&$ 1.20_{-0.02}^{+0.02} $\\
	$0.07 $&$ 0.530\pm 0.004 $&$ 1.28_{-0.01}^{+0.01} $\\
	$0.09  $&$ 0.600\pm 0.004 $&$ 1.35_{-0.1}^{+0.02} $\\
\hline\hline
\end{tabular}
\end{center}
\label{table:hadr}
\end{table}

\section{Chiral effective parameters in the IILM}\label{extrapolation}

 In the previous sections, we have shown that the IILM provides a realistic description of the microscopic dynamics responsible for chiral symmetry and that it contains pions as low-energy vacuum excitations. 
Thus, we conclude that the effective theory approximates the path integral sufficiently well that the IILM can be described by chiral perturbation theory, and we therefore ask the next question of how similar the low energy constants  are to those arising in QCD.  Hence, we determine the quark condensate and the pion decay constant from the dependence of the pion mass on the quark mass. Note that the numerical value of the quark mass in QCD depends on the renormalization scale. In the IILM we do not have this  freedom since  the ultraviolet cut-off scale is provided by the  inverse instanton size $1/\bar{\rho}\simeq600~\MeV$.
Chiral perturbation theory to $\mathcal{O}(p^4)$ predicts a dependence of  the form:
\be
m_{\pi}^2=2m_qB_0\Big(1+\frac{2m_qB_0}{32\pi^2f_0^2}\ln\Big(\frac{2mB_0}{\Lambda^2}\Big)\Big).
\ee
The chiral scale $\Lambda$ is set at the $\rho$ vector meson mass, which in this model is found to be independent on the quark mass, with the value $M_{\rho}=1$ GeV~\cite{IILMmeson}. Then, using the value for $\frac{B_0}{f_0^2}=340\ \GeV^{-1}$ which was extracted from the analysis of the Dirac spectrum (\ref{eq:b0ovf0}), one can  extract $B_0$. 

The results of the chiral fit to the numerical IILM calculations are shown in Fig. \ref{fig:pionfit}. 
\begin{figure}
	\centering
		\includegraphics[width=1.\textwidth]{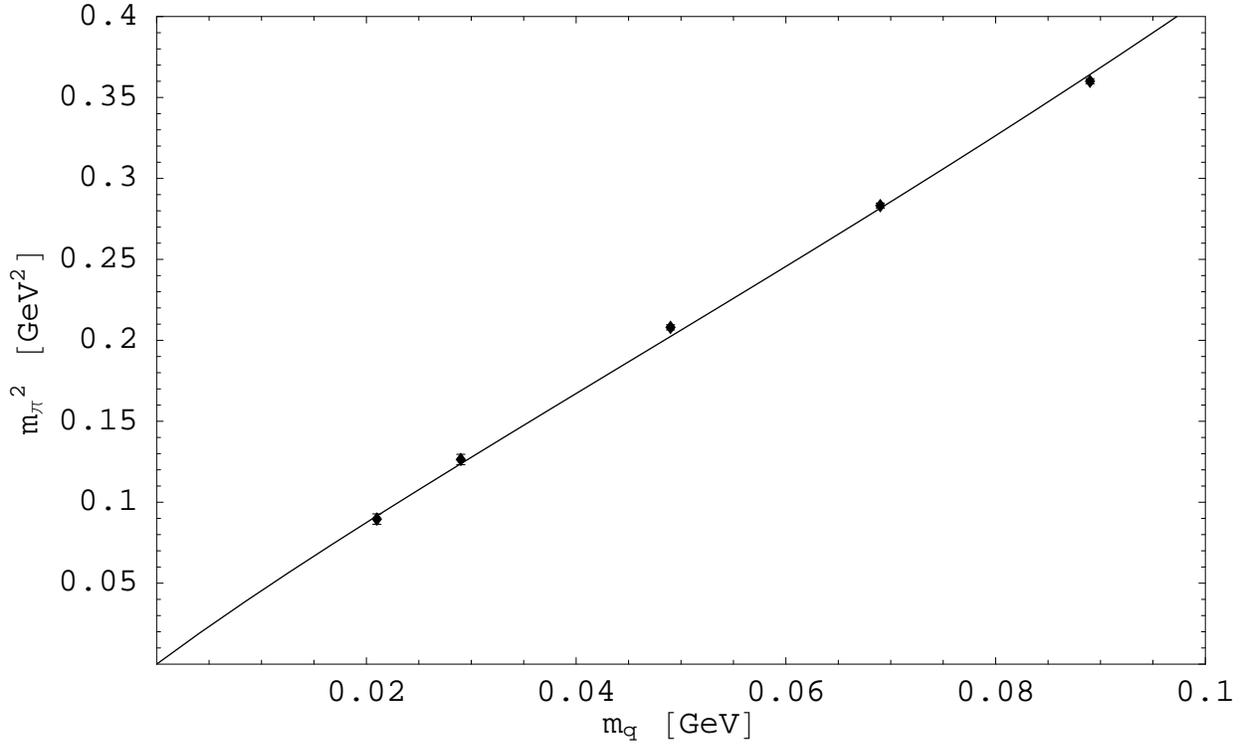}
	\caption{Chiral extrapolation of the pion mass, obtained in the IILM using $\mathcal{O}(p^4)$  $\chi$pt.}\label{fig:pionfit}
\end{figure}
The corresponding low-energy chiral coefficients calculated in our model are 
\be
f_0=0.085 \pm 0.003\ \GeV, \\
B_0=2.43 \pm 0.02\ \GeV, 
\ee
corresponding to a chiral condensate of 
\be
\langle \bar{q} q\rangle \approx  -(0.259\ \textrm{GeV})^3.
\ee
The fact that these quantities are rather close to the corresponding values extracted in QCD
 implies that the low-energy effective theory of the IILM is indeed not far from that of QCD.  Note that a recent analysis of the quark mass dependence of the chiral condensate in the instanton vacuum  leading to comparable results} can be found in \cite{musak:2006}. 

 A similar analysis can be carried out in the nucleon sector, where several approaches to chiral extrapolation have been proposed, for example see \cite{weise,procura,Detmold:2004ap,Leinweber:2005xz,Meissner:2004npa} and references therein. One can, eg., follow Procura {\it et al.}~\cite{weise,procura} and consider covariant baryon chiral perturbation theory (B$\chi$PT) with infrared regularization at order $\mathcal{O}(p^4)$~\cite{Becher:1999}. In this scheme the nucleon mass expansion in powers of the pion mass gives the following result:
\begin{eqnarray}\label{eq:nuclmassp4}
M_N&=&M_0-4c_1 m_\pi^2-\frac{3g_A^2}{64\pi f_\pi^2}m_\pi^3\nonumber\\
 &+& \Big[4e_1^r(\lambda)-\frac{3}{64\pi^2 f_\pi^2}\Big(\frac{g_A^2}{M_0}-\frac{c_2}{2}\Big)-\frac{3}{32\pi^2f_\pi^2}\Big(\frac{g_A^2}{M_0}-8c_1+c_2+4c_3\Big)\log\frac{m_\pi}{\Lambda}\Big]m_{\pi}^4\nonumber\\
 &+&\frac{3g_A^2}{256\pi f^2_\pi M_0^2}m^5_\pi
\end{eqnarray}
 where $e_1^r(\lambda)$ is the finite part of 
\begin{equation}
	e_1=e_1^r(\lambda)+\frac{3L}{2f_{\pi}^2}\Big(\frac{g_A^2}{M_0}-8c_1+c_2+4c_3\Big),
\end{equation}
and encodes information on the unresolved ultra-violet physics.

\begin{figure}
        \centering
                \includegraphics[width=1\textwidth]{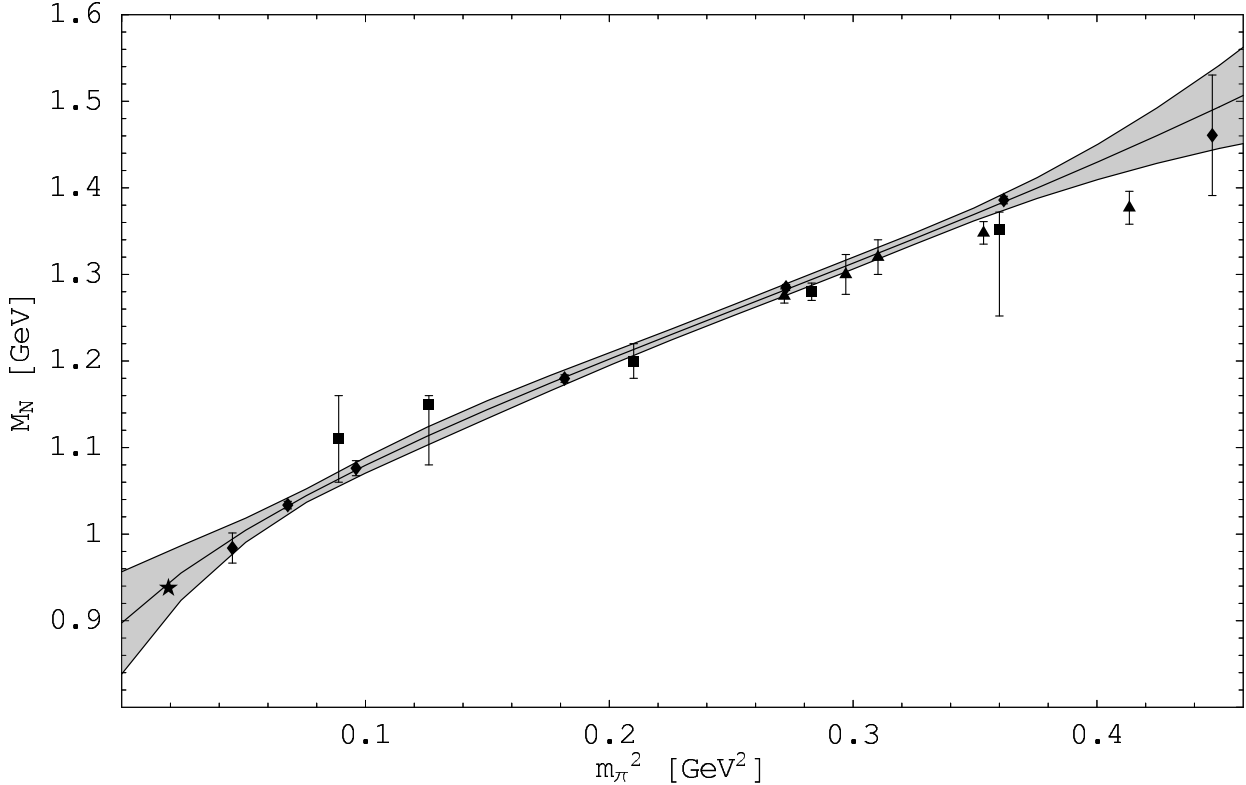}
        \caption{Nucleon mass chiral extrapolation at $\mathcal{O}(p^4)$ in relativistic B$\chi$pt. The error band is determined according to the method outlined  in \cite{procura}. The IILM, CP-PACS, and MILC data are denoted by the squares, triangles, and diamonds respectively, and the star represents the physical point.}\label{fig:fitmn}
\end{figure}

 We have performed a fit of the MILC and CP-PACS data in the region $m_\pi<0.6$ GeV, supplemented by the nucleon mass at the physical point. To reduce the number of free parameters, we fixed  $f_\pi$ and $g_A$ to their experimental values and $c_2=3.2~\textrm{GeV}^{-1}$, in accordance with the analysis of the $\pi N$ scattering in \cite{Fettes:1998}. 
The result of our fit is shown in Fig. \ref{fig:fitmn}, where it is compared with our predictions for the nucleon mass in the range $300~\textrm{MeV}\lesssim M_N\lesssim 600~\textrm{MeV}$.
 Although the present statistical and systematic errors in the calculated proton mass do not allow us to attempt a direct quantitative determination of  the effective parameters of B$\chi$pt for the IILM alone, the overall agreement between IILM and lattice data indicates  that they are consistent with those extracted from QCD at the level of one standard deviation. 

\section{Instanton-Induced Dynamics in the Chiral Regime}\label{chiraldynamics}

\begin{figure}
\includegraphics[width=0.9\textwidth]{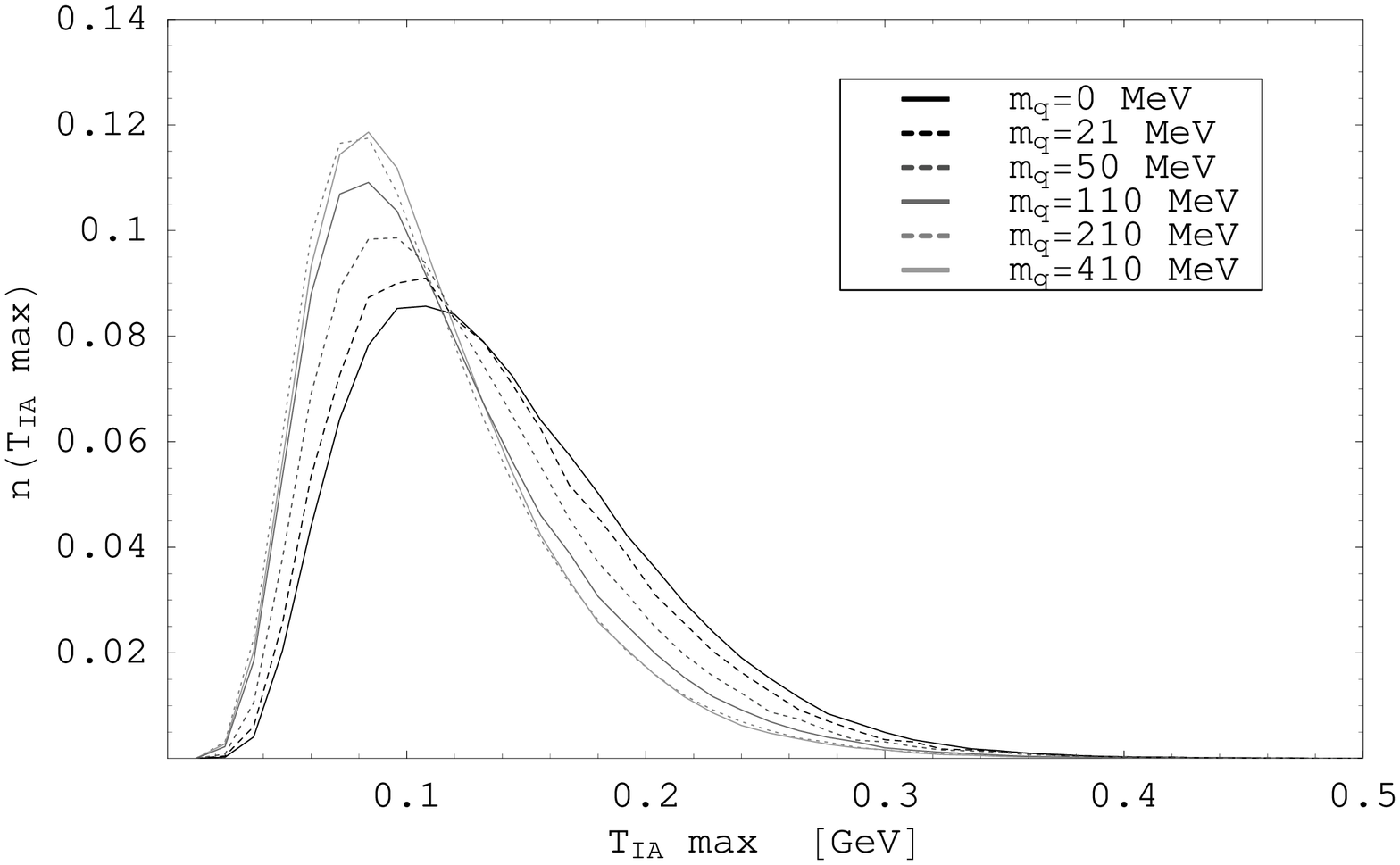}
\caption{Distribution of the maximum fermionic overlap matrix elements in the IILM for different quark masses.}
\label{fig:tia}
\end{figure}

 Having established that for $m_\pi\lesssim~500$~MeV the IILM agrees at a qualitative and even quantitative level with several QCD predictions obtained from $\chi$pt and lattice calculations, we can now use the physical insight from that model concerning the instanton-induced correlations to explore the dynamical mechanisms involved in the transition into the chiral regime.

\begin{figure}
	\centering
		\includegraphics[width=.9\textwidth]{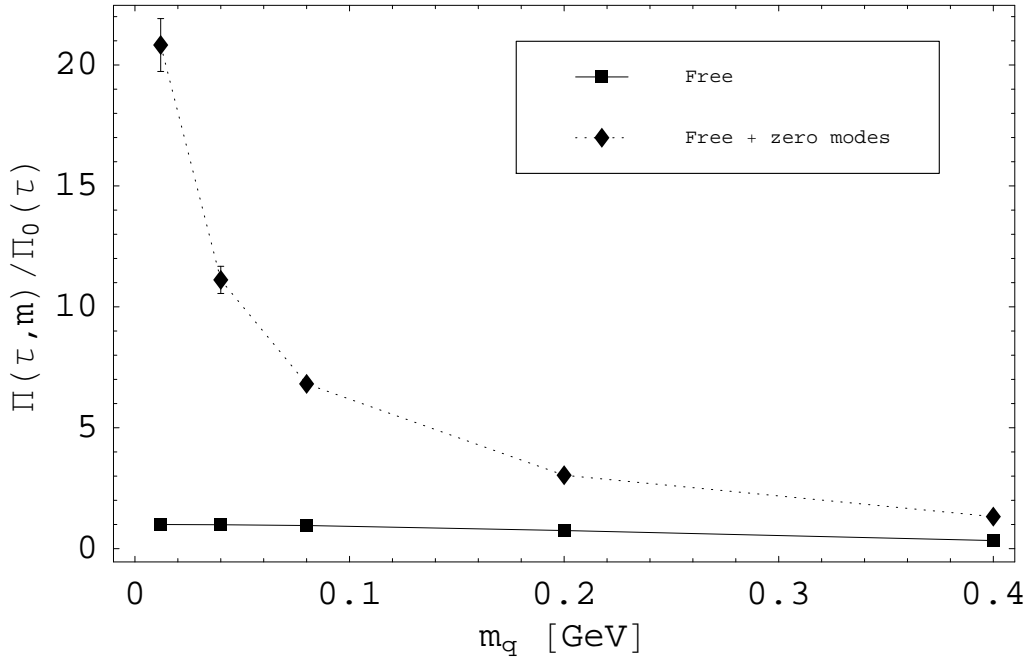}
	\caption{Sum of the zero-mode and free contributions to the pseudoscalar point-to-point correlation function computed in the IILM at $\tau=1~\fm$ for different quark masses  (diamonds) compared with the free contribution alone (squares).}
\label{pscompare}
\end{figure}

From (\ref{Szm}), it follows that the non-perturbative dynamics associated with  the instanton-induced near-zero modes becomes parametrically small for quark masses much larger than the typical value of the overlap matrix element $T_{I J}$.  
In Fig.\ref{fig:tia} we have plotted the distributions of the {\it maximum} overlap matrix elements $T_{IA}$ obtained at different values of the quark mass. We note that such distributions are peaked around the value $m^\star\simeq~80$~MeV. The value $m^\star$ plays a central role in specifying the scale for instanton induced chiral symmetry breaking in the IILM. Physically, we  expect that  for $m_q~\gg~m^\star$, corresponding to $m_\pi\gg 500~$MeV,  the interactions associated with chiral dynamics become sub-leading. We note that this number is consistent with the phenomenological estimate derived in \cite{sia}, using the single instanton approximation.
To further illustrate this transition,  in Fig.~\ref{pscompare} we show the point-to-point pseudoscalar correlation function evaluated in the IILM in the zero-mode approximation (zma) in which
\be
	S(x,y)=S_{free}(x,y)+S_{zm}(x,y). 
\ee

We normalize this full correlation function to the corresponding correlation computed in the free massless theory, 
\be
\frac{\Pi(\tau,m_q)}{\Pi_0(\tau)}&=&\frac{\langle 0| j_5(\tau) j^\dagger_5(0)
|0\rangle_{IILM-free+zm}}{4 \pi^4/\tau^6}\\
j_5(x)&=&\bar{u}(x) \gamma_5 d(x),
\ee
evaluated at a typical non-perturbative scale $\tau=1$~fm for different quark masses, and compare it to the free massless theory without zero mode contributions. 
We observe that for small quark masses, the contribution of the zero-mode part of the propagators completely saturates the propagator and is more than an order  of magnitude larger than the free contribution. On the other hand, as the quark mass increases,  the zero-mode contribution is gradually suppressed and for $m_q\simeq~200~\MeV$ it is only a few times larger than the free contribution. 

As the quark mass gets larger and larger, the quark loop contribution to QCD correlation functions becomes more and more suppressed. Interestingly, in the IILM, the scale $m^\star$ also determines the regime where  the quenched approximation is expected to become reliable. In fact, from Eq.s (\ref{det}),
(\ref{detzm}) and (\ref{detnzm}), we see that for quark masses $m_q\gg \textrm{max}[T_{IJ}]\simeq m^\star$, the contribution of the overlap matrix elements to the fermionic determinant becomes negligible. On the other hand, in this model, the mass contribution to the ultraviolet sector of the spectrum factorizes and therefore cancels out in all ensemble averages.

The numerical results presented in the previous section reveal two interesting features of the Instanton Model: (i) for $m_q\lesssim m^\star$ the density of quasi zero-modes increases as the quark mass decreases  and (ii) for $m_q\gtrsim m^\star$ the near zero-mode part of the spectrum becomes practically independent on the quark mass. These two non-trivial dynamical effects are in fact related and can be explained as follows.
Quark loops are known to generate strong non-local correlations between pseudoparticles of opposite topological charge. Such correlations tend to suppress configurations in which one or more pseudoparticles are located far from all others.  Hence, the contribution of the fermionic determinant leads to a  reduction of the density of {\it nearly} zero-modes, as completely isolated instantons are known to have {\it exact} zero-modes.  As the quark mass gets larger, such a topological screening becomes less and less effective and the population of near exact zero-modes increases, explaining  the rise of the peak of eigenvalue density near the origin. On the other hand, we have seen in section \ref{IILM} that for $m_q\gtrsim m^\star$, the contribution of the fermionic determinant is suppressed and fermion-induced topological screening completely disappears. As a result, the low-virtuality sector of the Dirac spectrum stops depending on the quark mass. These are  interesting  predictions of the instanton liquid model that can be checked with unquenched lattice calculations.

We conclude this section by emphasizing that the existence of a single mass scale $m^*$ at which chiral symmetry breaking begins to diminish and at which the quenched approximating begins to become valid is a characteristic prediction of the instanton model. In principle, in full QCD we expect the scales for such transitions to be implicit functions of the only dimensional parameters of the theory, $\Lambda_{QCD}$ and $m_q$ and there is no  {\it a priori} reason for which they should coincide, unless they are driven by a common dynamical mechanism.

\section{Conclusions}
\label{conclusions}

In this work we have used  the IILM to investigate the dynamical mechanisms that drive the dynamics and the structure of hadrons in the light quark sector of QCD. To this end, we have checked that the model's predictions are consistent with QCD in the pion mass regime $\lesssim 500~$MeV, where we expect chiral dynamics to play an important role.
By computing the nucleon mass and pion mass at different values of the quark mass we have shown that the IILM is  consistent with the existing lattice data in the range $m_{\pi}\sim 300-600$ MeV. 
We have observed that the best agreement with lattice QCD data in this region is obtained for a value of the average instanton size $\bar{\rho}=0.32~$fm, which is close to Shuryak's earlier phenomenological  estimate. On the other hand, the calculated instanton density  $n\approx 3~ \textrm{fm}^{-4}$ is considerably larger than in earlier phenomenological estimates and closer to  lattice results.
By studying the dependence on the quark mass of the density of eigenvalues of the Dirac operator, we have shown that the model also contains the correct dynamics to reproduce predictions of $\chi$pt in the small quark mass regime where such a theory is applicable and finite-order calculations are reliable. 
We have shown that, in the chiral limit, our simulations converge to the $N_f=2$ prediction: $\lim_{\lambda\to 0} \rho'(\lambda)=0$. 
We have studied the modification of the Dirac spectrum induced by  small quark masses and found that they are  proportional to $ m_q^{\alpha}\lambda^{1-\alpha}$, a functional form that generates a $1/m_q$ power-law divergence in the quark mass dependence of the scalar three-point correlators as predicted by  $\chi$pt~\cite{smilga}.   We have discussed how the qualitative structure of such mass corrections has a simple physical interpretation in terms of quark-loop-induced instanton-antiinstanton correlations. It would be very interesting to check if the corrections predicted in the IILM are observed in full lattice QCD simulations.

From a chiral extrapolation of the pion mass,  we  computed the effective chiral parameters $f_\pi$ and $B$ and showed that they are comparable to those in QCD.  Similarly, from the agreement of our results for the nucleon mass at different quark masses with the available lattice simulations,  we conclude that the parameters for the baryon chiral perturbation theory are consistent with those extracted from QCD, although our present statistics do not allow us to perform a direct chiral extrapolation.

Having checked that chiral dynamics is correctly incorporated in the instanton model, we have exploited our analytic understanding of instanton-induced correlations  to study the dynamical mechanisms involved in the transition into the chiral regime at a microscopic level. We have identified a mass scale  $m^\star=80$~MeV above which we do not expect QCD correlators to be dominated by chiral dynamics and above which quenched calculations should begin to approximate full QCD.  

 If the computational technology is developed to the point that the present
 statistical and systematic errors on the nucleon mass are significantly improved, this framework could represent a complementary tool to perform direct chiral extrapolation of a variety of observables of interest, such as moments of DIS and generalized parton distributions, magnetic moments and form factors. 
Another interesting development would be to investigate if unstable light-quark resonances that receive contribution at order $n^2$ level in the instanton density, such as the $\rho$-meson and $\Delta$-isobar,   can also be well described by the 't~Hooft interaction. In this context it would also be interesting to investigate the effect of confinement on the hadron masses, extending the pseudoparticle ensemble to include both regular and singular gauge instantons~\cite{Lenz:2003jp}.

\acknowledgements
We thank C. Gattringer, F. Lenz, M. Procura and E. V. Shuryak for useful discussions and
the Institute for Nuclear Theory at the University of Washington for its hospitality. This work was supported in part by the U.S. DOE office of Nuclear Physics under contract DE-FC02-94ER40818 and the INFN-MIT "Bruno Rossi" Exchange Program. 
\bibliography{ILM}
\end{document}